\documentstyle[epsfig,12pt]{article}
\newlength{\dinwidth}
\newlength{\dinmargin}
\newcommand{\epsl}{\varepsilon \hspace{-5pt} / }
\newcommand{\rsl}{r \hspace{-5pt} / }
\newcommand{\qsl}{q \hspace{-5pt} / }
\newcommand{\pbsl}{p \hspace{-5pt} / }
\newcommand{\pssl}{p' \hspace{-7pt} / }

\newcommand{\ra}{\rightarrow}
\newcommand{\BGAMAXS}{B \ra X _{s} + \gamma}

\newcommand{\ba}{\begin{array}}
\newcommand{\ea}{\end{array}}
\newcommand{\be}{\begin{equation}}
\newcommand{\ee}{\end{equation}}
\newcommand{\bea}{\begin{eqnarray}}
\newcommand{\eea}{\end{eqnarray}}

\def\bra{\langle}
\def\ket{\rangle}

\def\a{\alpha}
\def\b{\beta}
\def\g{\gamma}
\def\d{\delta}
\def\e{\epsilon}
\def\p{\pi}

\def\ep{\varepsilon}

\def\l{\lambda}
\def\m{\mu}
\def\n{\nu}
\def\G{\Gamma}

\def\to{\rightarrow}
\begin{document}
\thispagestyle{empty}
\addtocounter{page}{-1}
\begin{flushright}
SLAC-PUB-7144\\
ZU-TH 7/1996\\
hep-ph/9603404\\
March 1996
\end{flushright}
\vspace*{1.8cm}
\centerline{\Large\bf Virtual $O(\a_s)$  corrections to the inclusive
decay  
$b \to s \gamma$ 
\footnote{Work supported in part by Schweizerischer
Nationalfonds and the Department of Energy, contract
DE-AC03-76SF00515}}
\vspace*{2.0cm}
\centerline{\large\bf Christoph Greub }
\vspace*{0.4cm}
\centerline{\large\it Stanford Linear Accelerator Center}
\centerline{\large\it Stanford University, 
Stanford, California 94309, USA}
\vspace*{0.8cm}
\centerline{\large\bf Tobias Hurth \footnote{ address after March 1996:
ITP, SUNY at Stony Brook, Stony Brook NY 11794-3840, USA} and Daniel Wyler}
\vspace*{0.4cm}
\centerline{\large\it Institute for Theoretical Physics, University of 
Z\"urich}
\centerline{\large\it Winterthurerstr. 190, CH-8057 Z\"urich, Switzerland}
\vspace*{2.5cm}
\centerline{\Large\bf Abstract}
We present in detail the calculation of the $O(\a_s)$ virtual corrections
to the matrix element for $b \to s \g$. Besides the
one-loop virtual corrections of the electromagnetic and color
dipole operators $O_7$ and $O_8$, we include the 
important two-loop
contribution of the four-Fermi operator $O_2$. By applying
the Mellin-Barnes representation to certain internal 
propagators, the
result of the two-loop diagrams is obtained analytically
as an expansion in $m_c/m_b$.
These results are then combined with existing $O(\a_s)$ Bremsstrahlung
corrections in order to obtain the inclusive rate for $B \to X_s \g$.
The new contributions drastically reduce the large renormalization
scale dependence of the leading logarithmic result. Thus a
very precise Standard Model prediction for this inclusive process
will become possible once also the corrections to the Wilson
coefficients are available.

\vspace*{1.5cm}
\centerline{Submitted to Physical Review D}

\newpage
\section{Introduction}
\label{sec:introd}
\setcounter{equation}{0}

In the Standard model (SM), flavor-changing neutral currents only 
arise at the one-loop level. 
This is why the corresponding rare B meson decays
are particularly sensitive to ``new physics''.  
However, even 
within the Standard model framework, one can use them 
to constrain the Cabibbo-Kobayashi-Maskawa matrix
elements which involve the top-quark. For both these
reasons,
precise experimental and theoretical work  on these
decays is required.

In 1993, $B \to K^* \gamma$ was the first rare B decay mode
measured
by the CLEO collaboration \cite{CLEOrare1}. Recently, also the first 
measurement of
the inclusive photon energy
spectrum and the branching ratio
in the decay $\BGAMAXS$ was reported \cite{CLEOrare2}.
In contrast to the exclusive channels, the inclusive mode allows a less
model-dependent comparison with theory, because no specific
bound state model is needed for the final state. This opens 
the road to a rigorous comparison with theory.

The data agrees with
the SM-based theoretical computations presented
in \cite{agalt,aglett,shifmangamma},
given that there are large uncertainties in both the
experimental and the theoretical results.
In particular, the measured branching ratio $BR(B \to X_s \gamma)
= (2.32 \pm 0.67)
\times 10^{-4}$ \cite{CLEOrare2} overlaps with the
SM-based estimates in
\cite{agalt,aglett} and in \cite{Buras94,Ciuchini94}.

In view of the expected increase in the experimental precision, 
the calculations must be refined correspondingly in order to
allow quantitative statements about new
physics or standard model
parameters. So far, only the leading logarithmic
corrections have been worked out systematically.
In this paper we evaluate an important class of next
order corrections, which we will describe in  detail below
\footnote{Some of the diagrams were calculated by Soares
\cite{Soares2}}.

We start within the usual framework of an effective theory with
five quarks, obtained by integrating out the
heavier degrees of freedom which  
in the standard model are the top quark and the $W$-boson. 
The effective Hamiltonian includes
a complete set of dimension-6 operators relevant for the process
$b \to s \gamma$  (and $b \to s \g g$) \cite{Grinstein90} 
\begin{equation}
\label{heff}
H_{eff}(b \to s \gamma)
       = - \frac{4 G_{F}}{\sqrt{2}} \, \lambda_{t} \, \sum_{j=1}^{8}
C_{j}(\mu) \, O_j(\mu) \quad ,
\end{equation}
with
$G_F$ being the Fermi
coupling constant and
$C_{j}(\mu) $ being the Wilson coefficients evaluated at the scale $\mu$,
and $\lambda_t=V_{tb}V_{ts}^*$ with $V_{ij}$ being the
CKM matrix elements.
The operators $O_j$ are as follows:
\bea
\label{operators}
O_1 &=& \left( \bar{c}_{L \b} \g^\m b_{L \a} \right) \,
        \left( \bar{s}_{L \a} \g_\m c_{L \b} \right)\,, \nonumber \\
O_2 &=& \left( \bar{c}_{L \a} \g^\m b_{L \a} \right) \,
        \left( \bar{s}_{L \b} \g_\m c_{L \b} \right) \,,\nonumber \\
O_3 &=& \left( \bar{s}_{L \a} \g^\m b_{L \a} \right) \, \left[
        \left( \bar{u}_{L \b} \g_\m u_{L \b} \right) + ... +
        \left( \bar{b}_{L \b} \g_\m b_{L \b} \right) \right] \,,
        \nonumber \\
O_4 &=& \left( \bar{s}_{L \a} \g^\m b_{L \b} \right) \, \left[
        \left( \bar{u}_{L \b} \g_\m u_{L \a} \right) + ... +
        \left( \bar{b}_{L \b} \g_\m b_{L \a} \right) \right] \,,
        \nonumber \\
O_5 &=& \left( \bar{s}_{L \a} \g^\m b_{L \a} \right) \, \left[
        \left( \bar{u}_{R \b} \g_\m u_{R \b} \right) + ... +
        \left( \bar{b}_{R \b} \g_\m b_{R \b} \right) \right] \,,
        \nonumber \\
O_6 &=& \left( \bar{s}_{L \a} \g^\m b_{L \b} \right) \, \left[
        \left( \bar{u}_{R \b} \g_\m u_{R \a} \right) + ... +
        \left( \bar{b}_{R \b} \g_\m b_{R \a} \right) \right] \,,
        \nonumber \\
O_7 &=& (e/16\p^{2}) \, \bar{s}_{\a} \, \sigma^{\m \n}
      \, (m_{b}(\mu)  R + m_{s}(\mu)  L) \, b_{\a} \ F_{\m \n} \,,
        \nonumber \\
O_8 &=& (g_s/16\p^{2}) \, \bar{s}_{\a} \, \sigma^{\m \n}
      \, (m_{b}(\mu)  R + m_{s}(\mu)  L) \, (\l^A_{\a \b}/2) \,b_{\b}
      \ G^A_{\m \n} \quad .
        \nonumber \\
\eea
 In the dipole type
operators $O_7$ and $O_8$, 
$e$ and $F_{\m \n}$ ($g_s$ and $G^A_{\m \n}$)
denote the electromagnetic (strong)
coupling constant and 
field strength
tensor, respectively. $L=(1-\g_5)/2$ and $R=(1+\g_5)/2$
stand for the left and right-handed projection operators.
It should be stressed in this context that the explicit 
mass factors in $O_7$
and $O_8$ are the running quark masses.

QCD corrections to the decay rate for $b \to s \g$
bring in
large logarithms of the form $\a_s^n(m_W) \, \log^m(m_b/M)$,
where $M=m_t$ or $m_W$ and $m \le n$ (with $n=0,1,2,...$).
One can systematically resum these large terms by renormalization
group techniques. Usually, one matches the full standard model theory
with the effective theory at the scale $m_W$. At this scale,
the large logarithms generated by matrix elements in the 
effective theory are the  same ones
as in the full theory. Consequently, the
Wilson coefficients only contain small QCD corrections.
Using the renormalization group equation, the Wilson coefficients
are then calculated at the scale $\mu \approx m_b$, 
the relevant scale for a $B$ meson decay.
At this scale
the large logarithms are contained in 
the Wilson coefficients
while the matrix elements of the operators are free
of them.

As noted, so far the decay rate for $b \to s \gamma$ has been 
systematically
calculated only to leading logarithmic accuracy i.e., $m=n$.
To this precision it is consistent to perform the 'matching' of the
effective and full theory
without taking into account QCD-corrections
\cite{Inami}
and to calculate
the anomalous dimension matrix to order $\a_s$ \cite{Ciuchini}.
The corresponding leading logarithmic Wilson coefficients
are given explicitly in
\cite{Buras94,AGM94}.
Their numerical values in the naive dimensional scheme (NDR) are 
listed in table 1 for different values of the renormalization
scale $\mu$. The leading logarithmic contribution to the
decay matrix element is then obtained by calculating the
tree-level matrix element of the operator $C_7 O_7$ and
the one-loop matrix elements of the four-Fermi operators
$C_i O_i$ ($i=1,...,6$).
In the NDR scheme
the latter
can be absorbed into a redefinition of $C_7 \to C_7^{eff}$
\footnote{For the analogous $b \to s g$ transition, the effects
of the four-Fermi operators can be absorbed by the shift
$C_8 \to C_8^{eff}=C_8 + C_5$.}
\be
\label{C78eff}
C_7^{eff} \equiv  C_7 + Q_d \, C_5 + 3 Q_d \, C_6 \quad .
\ee
In the `t Hooft-Veltman scheme (HV) \cite{HV},
the contribution of the four-Fermi
operators vanishes. The Wilson coefficients
$C_7$ and $C_8$ in the HV scheme
are identical to $C_7^{eff}$ and $C_8^{eff}$ in the
NDR scheme.
Consequently, the complete leading logarithmic result
for the decay amplitude $b \to s \g$ is
indeed scheme independent.

Since the first order calculations have 
large
scale uncertainties,
it is important to take into account the next-to-leading
order corrections. They are most prominent in the
photon energy spectrum. While it is a
$\delta$-function
(which is smeared out by the Fermi motion of the $b$-quark inside
the $B$ meson) in the leading order, Bremsstrahlung corrections, i.e.
the process $b \to s \gamma g$, broaden the shape of the spectrum
substantially. Therefore, these
important corrections
have been taken into account
for the contributions of the operators $O_7$ and $O_2$ some time
ago \cite{agalt} and recently
also of the full
operator basis \cite{aglett,aglong,Pott}.
As expected,
the contributions of $O_7$
and $O_2$ are by far the most important ones, especially
in the experimentally accessible part of the spectrum.
Also those
(next-to-leading) corrections,
which are necessary to cancel the
infrared
(and collinear)  singularities of
the Bremsstrahlung diagrams  were 
included.
These are the virtual gluon corrections to the contribution
of the operator $O_7$ for $b \to s \g$ and the virtual
photon corrections to $O_8$ for $b \to s g$.

A complete next-to-leading calculation implies
two classes of improvements: First,
the Wilson coefficients to next-leading
order  at the scale $\mu \approx m_b$ are required. To this end
the matching
with the full theory (at $\mu=m_W$)
must be done at the $O(\a_s)$ level and the renormalization
group equation has to be solved using the anomalous dimension
matrix calculated up to order $\a_s^2$.
Second,
the virtual $O(\a_s)$ corrections for the matrix element (at scale
$\mu \approx m_b$)
must be evaluated and combined with the Bremsstrahlung corrections.
The higher order matching has been calculated in ref. \cite{adel}
and work on the Wilson coefficients is in progress.
In this paper we will evaluate all the virtual
correction beyond those evaluated already in connection with the
Bremsstrahlung process. We expect them
to reduce substantially
the strong scale dependence of the leading order calculation.

Among the four-Fermi operators only $O_2$
contributes sizeably and we calculate only
its virtual corrections
to the matrix element for $b \to s \g$.
The matrix element $O_1$ vanishes
because of color,
\begin{table}[htb]
\label{coeff}
\begin{center}
\begin{tabular}{| c | c | r | r | r | }
\hline
 $C_i(\mu)$ & $\mu=m_W$ & $\mu=10.0$ GeV
                              & $\mu=5.0$ GeV
                              & $\mu=2.5$ GeV\\
\hline \hline
$C_1$ & $0.0$ & $-0.149$ & $-0.218$ & $-0.305$ \\
$C_2$ & $1.0$ & $1.059$ & $1.092$ & $1.138$ \\
$C_3$ & $0.0$ & $0.006$ & $0.010$ & $0.014$ \\
$C_4$ & $0.0$ & $-0.016$ & $-0.023$ & $-0.031$ \\
$C_5$ & $0.0$ & $0.005$ & $0.007$ & $0.009$ \\
$C_6$ & $0.0$ & $-0.018$ & $-0.027$ & $-0.040$ \\
$C_7$ & $-0.192$ & $-0.285$ & $-0.324$ & $-0.371$ \\
$C_8$ & $-0.096$ & $-0.136$ & $-0.150$ & $-0.166$ \\
$C_7^{eff}$ & $-0.192$ & $-0.268$ & $-0.299$ & $-0.334$ \\
$C_8^{eff}$ & $-0.096$ & $-0.131$ & $-0.143$ & $-0.157$ \\
\hline
\end{tabular}
\end{center}
\caption{Leading logarithmic Wilson coefficients $C_i(\mu )$
at the matching scale $\mu=m_W=80.33$ GeV
and at three other scales, $\mu = 10.0$ GeV,
$\mu =5.0$ GeV and $\mu = 2.5$ GeV. For $\a_s(\mu)$
(in the $\overline{MS}$ scheme) we used the
one-loop expression with 5 flavors and $\a_s(m_Z)=0.117$.
The entries correspond to the top
quark mass
 $\overline{m_t}(m_{t,pole})=170$ GeV (equivalent to
$m_{t,pole}= 180 $ GeV).}
\label{table1}
\end{table}
and the
penguin induced four-Fermi operators $O_3,...., O_6$ can be
neglected \footnote{
This omission will be a source of a slight scheme and scale dependence
of the next-to-leading order result.}
because their Wilson
coefficients
\footnote{It is consistent to calculate the corrections
using the leading logarithmic Wilson coefficients.}
are much smaller than $C_2$, as illustrated
in table \ref{coeff}.
However, we do take into account the virtual $O(\a_s)$
corrections to $b \to s \g$ associated with the
magnetic operators $O_7$ (which has already been calculated in the
literature) and $O_8$ (which is new).
Since the corrections to $O_7$ and $O_8$ are one-loop diagrams,
they are relatively easy to work out.
In contrast, the corrections to $O_2$, involve two-loop
diagrams, since this operator itself only contributes
at the one-loop level.

Since the virtual and Bremsstrahlung
corrections to the matrix elements are only one (well-defined)
part of the whole next-to-leading program,
we expect that this contribution alone will depend on
the renormalization scheme used. Even within the modified minimal
subtraction scheme $(\overline{MS})$ used here,
we expect that two different
``prescriptions'' how to treat $\gamma_5$, will lead to different
answers. Since previous calculations of the Bremsstrahlung
diagrams have been done in the NDR scheme and also the leading
logarithmic Wilson coefficients are available in this scheme,
we also use it here.
For future checks, however, we also consider in Appendix A the
corresponding calculation in the HV scheme.

The remainder of this paper
is organized as follows. In section 2 we give the
two-loop corrections for $b \to s \g$ based on the operator $O_2$
together with the counterterm contributions.
In section 3  the virtual corrections for $b \to s \g$
based on $O_7$ are reviewed including 
some of the Bremsstrahlung corrections.
Then, in section 4 we calculate the one-loop
corrections to $b \to s \g$
associated with $O_8$. Section 5 contains the results
for the branching ratio for $b \to s \g (g)$ and especially the 
drastic reduction of the renormalization scale dependence
due to the new contributions.
Appendix A contains the result of the
$O_2$ two-loop calculation in the HV scheme and, finally, 
to make the paper self-contained, we include in Appendix B
the Bremsstrahlung corrections to the operators $O_2$, $O_7$
and $O_8$. 

\section{Virtual corrections to $O_2$ in the NDR scheme}
\setcounter{equation}{0}
In this section
we present the calculation of the matrix element of the operator $O_2$
for $b \to s \g$ up to order $\a_s$ in the NDR scheme. 
The one-loop
$(\a_s^0)$  matrix element
vanishes and we must consider
several two-loop contributions.
Since they involve ultraviolet singularities also
counterterm contributions are needed. These are
easy to obtain, because  the operator renormalization constants
$Z_{ij}$ are known with enough
accuracy from the order $\alpha_s$
anomalous dimension matrix \cite{Ciuchini}.
Explicitly,
we need the contributions of the operators
$C_2 \delta Z_{2j} O_j$ to the matrix element for
$b \to s \gamma$,
where $\delta Z_{2j}$ denote the order $\alpha_s$
contribution of the operator renormalization constants.
In the NDR scheme, the non-vanishing
counterterms come from the one-loop matrix
element of $C_2 \delta Z_{25} O_5$ and
$C_2 \delta Z_{26} O_6$ as well as from the tree level
matrix element of the operator $C_2 \delta Z_{27} O_7$.
We also note that there are no contributions to $b \to s \g$
from counterterms proportional to evanescent 
operators multiplying the Wilson coefficient $C_2$.

\subsection{Regularized two-loop contribution of $O_2$}
\begin{figure}[htb]
\vspace{0.10in}
\centerline{
\epsfig{file=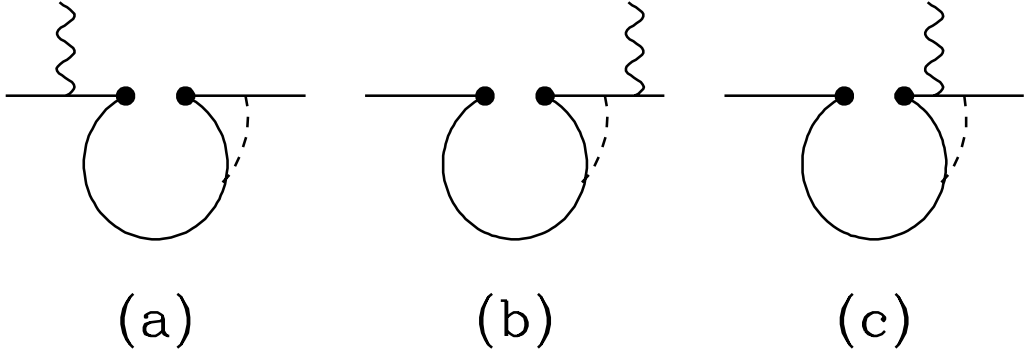,height=2in,angle=0,clip=}
}
\vspace{0.08in}
\caption[]{Diagrams 1a, 1b and 1c associated with the operator $O_2$.
The fermions ($b$, $s$ and $c$ quark) are represented by solid lines.
The wavy (dashed) line represents the photon (gluon).
\label{fig:1}}
\end{figure}

\begin{figure}[htb]
\vspace{0.10in}
\centerline{
\epsfig{file=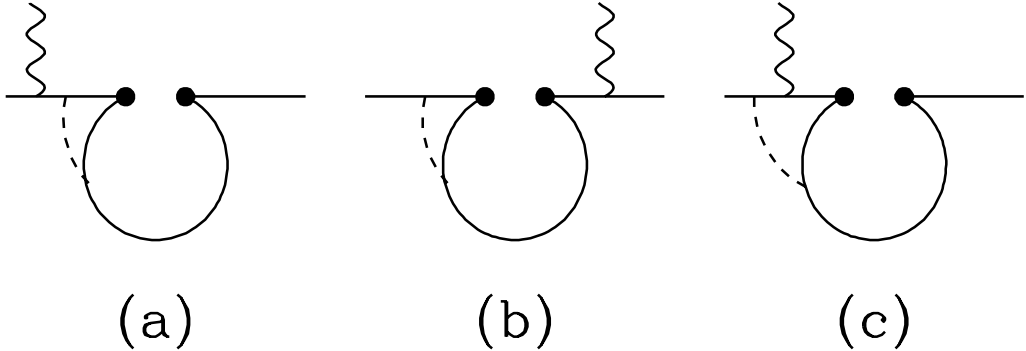,height=2in,angle=0,clip=}
}
\vspace{0.08in}
\caption[]{Diagrams 2a, 2b and 2c associated with the operator $O_2$.
\label{fig:2}}
\end{figure}

\begin{figure}[htb]
\vspace{0.10in}
\centerline{
\epsfig{file=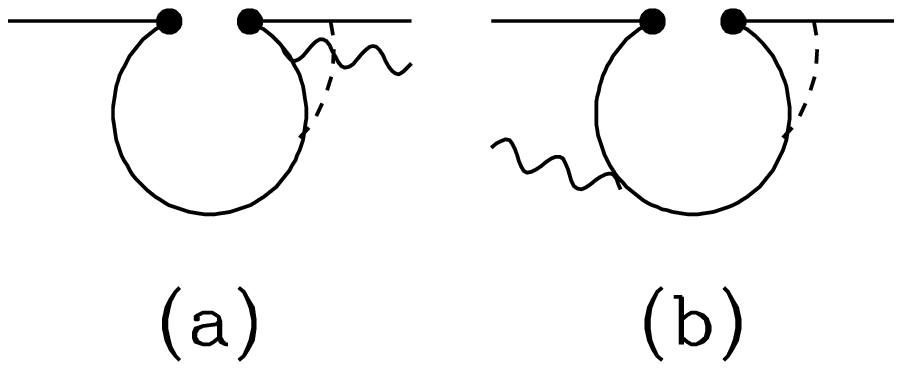,height=1.5in,angle=0,clip=}
}
\vspace{0.08in}
\caption[]{Diagrams 3a and 3b associated with the operator $O_2$.
We calculate directly their sum and denote it by $M_2(3)$, see text.
 \label{fig:3}}
\end{figure}

\begin{figure}[htb]
\vspace{0.10in}
\centerline{
\epsfig{file=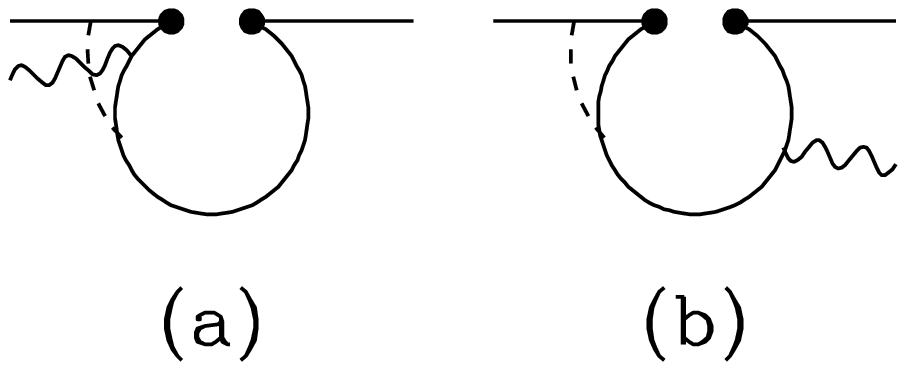,height=1.5in,angle=0,clip=}
}
\vspace{0.08in}
\caption[]{Diagrams 4a and 4b associated with the operator $O_2$.
We calculate directly their sum and denote it by $M_2(4)$, see text.
\label{fig:4}}
\end{figure}

The dimensionally regularized matrix element $M_2$
of the operator $O_2$ for $b \to s \g$
\be
\label{defm2}
M_2 = \bra s \gamma|O_2|b \ket
\ee
can be divided
 into 4 classes 
of non-vanishing two-loop
diagrams, as shown in Figs.
\ref{fig:1}--\ref{fig:4}.
The sum of the diagrams
in each class (=figure) is gauge invariant.
The contributions to the matrix element $M_2$
of the individual classes \ref{fig:1}--\ref{fig:4}
are denoted by
$M_2(1), M_2(2), M_2(3)$ and $M_2(4)$, where e.g.
$M_2(1)$
is
\be
\label{eich}
M_2(1) \equiv M_2(1a) + M_2(1b) + M_2(1c) \quad .
\ee

\begin{figure}[htb]
\vspace{0.10in}
\centerline{
\epsfig{file=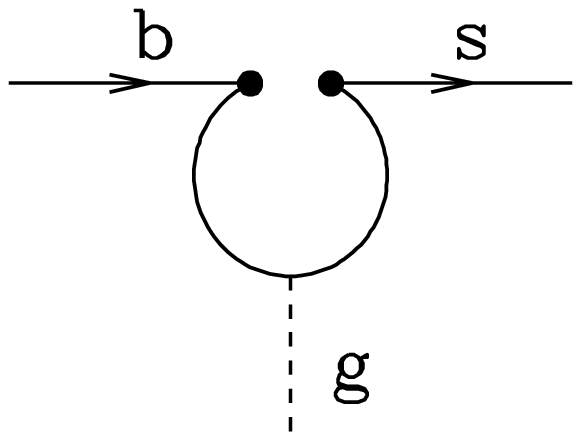,height=1.5in,angle=0,clip=}
}
\vspace{0.08in}
\caption[]{Building block $I_\b$ for the diagrams in Figs. (\ref{fig:1})
and (\ref{fig:2}) with an off-shell gluon.
\label{fig:5}}
\end{figure}

\begin{figure}[htb]
\vspace{0.10in}
\centerline{
\epsfig{file=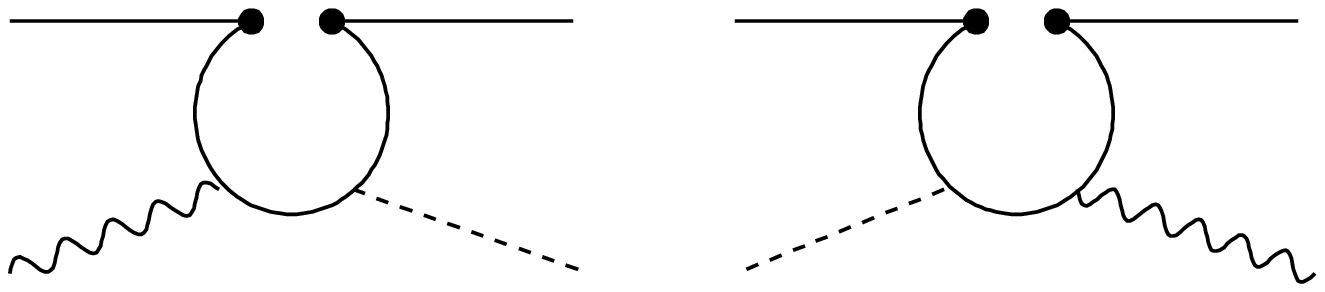,height=1.5in,angle=0,clip=}
}
\vspace{0.08in}
\caption[]{Building block $J_{\a\b}$
for the diagrams in Figs. (\ref{fig:3})
and (\ref{fig:4}).
\label{fig:6}}
\end{figure}

The main steps of the calculation are the following: We first
calculate the Fermion loops in the individual diagrams,
i.e.,  the 'building blocks' shown in
Fig. \ref{fig:5} and in Fig. \ref{fig:6},
combining together the
two diagrams in Fig. \ref{fig:6}.
As usual, we work in $d=4-2 \e$
dimensions; the results are presented as integrals over
Feynman parameters after
integrating over the (shifted) loop-momentum.
Then we insert these
building blocks into the full two-loop diagrams.
Using the Feynman parametrization again,
we calculate the integral over the second loop-momentum.
As the remaining Feynman parameter integrals contain rather
complicated denominators, we do not evaluate them directly.
At this level we also do not expand in the regulator $\e$.
The heart of our procedure
which will be explained more explicitly below,
is to
represent these denominators as  complex Mellin-Barnes integrals
 \cite{Abra}. After inserting this
representation and interchanging the order of integration, the
Feynman parameter integrals are reduced to well-known Euler
Beta-functions. Finally, the residue theorem
allows  to write the result of the remaining
complex integal as the sum over the residues taken at
the pole positions of
certain Beta- and
Gamma-functions; this naturally leads
to an expansion in the ratio $z=(m_c/m_b)^2$, which
numerically is about $z=0.1$.

We express 
the diagram in Fig. \ref{fig:5} (denoted by $I_\b$)
in a way convenient for
inserting into the two-loop
diagrams. As we will use $\overline{MS}$ subtraction later on,
we introduce the renormalization scale in the form
$\mu^2 \exp(\gamma_E) /(4 \pi)$, where
$\g_E=0.577...$ is the Euler constant.
Then,
$\overline{MS}$ corresponds to subtracting
the poles in $\e$.
In the NDR scheme, $I_\beta$ is given by
\footnote{The fermion/gluon and the fermion/photon couplings
are defined according to the covariant derivative
$D = \partial + i g_s \frac{\lambda^B}{2} A^B + i e Q A$}
\bea
\label{build1}
I_\beta &=& - \frac{g_s}{4 \pi^2} \, \Gamma(\e) \,
\mu^{2 \e} \,
\exp(\gamma_E \e)
\, (1-\e)
\exp(i \pi \e) \, \left(r_\b \rsl - r^2 \g_\b \right) \, L
\frac{\lambda}{2} \times \nonumber \\
&& \int_0^1 [x(1-x)]^{1-\e} \,
\left[ r^2 - \frac{m_c^2}{x(1-x)} + i \delta \right]^{-\e}
\quad ,
\eea
where $r$ is the four-momentum of the (off-shell) gluon,
$m_c$ is the mass of the charm quark propagating in the loop
and the term $i \delta$ is the "$\e$-prescription".
The free index $\b$ will be contracted with
the gluon propagator when inserting the building block into
the two-loop diagrams in Figs. \ref{fig:1} and \ref{fig:2}.
Note that $I_\beta$ is gauge invariant
in the sense that $r^\beta I_\beta = 0$.

Next we give the sum of the two diagrams in Fig. \ref{fig:6}, using
the decomposition in \cite{Simma_Wyler}.
We get (denoting this building block by $J_{\a\b}$)
\bea
\label{build2}
J_{\a\b}  &=& \frac{e g_s Q_u}{16 \pi^2} \,
\left[  E(\a,\b,r)  \, \Delta i_5
       + E(\a,\b,q) \, \Delta i_6
       - E(\b,r,q) \, \frac{r_\a}{(qr)} \Delta i_{23}
       \nonumber \right. \\ && \left. \hspace{1.0cm}
       - E(\a,r,q) \, \frac{r_\b}{(qr)} \Delta i_{25}
       - E(\a,r,q) \, \frac{q_\b}{(qr)} \Delta i_{26}
       \right] L \frac{\lambda}{2} \quad ,
\eea
where $q$ is the four-momentum
of the photon. The index $\a$ in eq. (\ref{build2})
is understood to be contracted with the polarization vector
$\ep$ of the photon, while the index $\b$ is contracted
with the gluon propagator in the  two-loop diagrams in Figs.
\ref{fig:3} and \ref{fig:4}.
The matrix $E$  in eq.
(\ref{build2}) is defined as
\be
\label{epsilongeneralization}
E(\a,\b,r) = \g_\a \g_\b \rsl - \g_\a r_\b + \g_\b (r_\a)
- \rsl g_{\a\b} .
\ee
In a four-dimensional
context these $E$ quantities can be reduced to expressions
involving the
Levi-Civit\`a tensor, i.e., $E(\a,\b,\g) = -i \,
\ep_{\a \b \g \mu} \, \g^\mu \g_5 $ (in the Bjorken-Drell
convention).
The dimensionally regularized expressions for the
$\Delta i$ read
\bea
\label{deltai5}
\Delta i_5 &=& -4 \int_S \, dx \, dy
\, \left[
4 (qr) x^2 y \e - 4 (qr) x y \e - 2 r^2 x^3 \e + 3 r^2 x^2 \e
\right. \nonumber \\ && \hspace{0.5cm} \left.
- r^2 x \e + 3 x C - C \right] \,
 [(1+\e)                                    \,
\Gamma(\e) \exp(\gamma_E \e) \mu^{2\e} \, C^{-1-\e}]
\eea
\bea
\label{deltai6}
\Delta i_6 &=& 4 \int_S \, dx \, dy
\, \left[
4 (qr) x y^2 \e - 4 (qr) x y \e - 2 r^2 x^2 y \e + 2 r^2 x^2 \e
+ r^2 x y \e
\right. \nonumber \\ && \hspace{0.5cm} \left.
- 2 r^2 x \e + 3 y C - C \right] \,
[(1+\e)
\Gamma(\e) \exp(\gamma_E \e) \mu^{2\e}  \, C^{-1-\e}]
\eea
\be
\label{deltai23}
\Delta i_{23} = -\Delta i_{26} = 8 (qr) \int_S \, dx \, dy \,
[x y \e \,
(1+\e)
\Gamma(\e) \exp(\gamma_E \e) \mu^{2\e} \, C^{-1-\e}]
\ee
\be
\label{deltai25}
\Delta i_{25} = -8 (qr) \int_S \, dx \, dy \,
[x(1-x)  \e \,
(1+\e)
\Gamma(\e) \exp(\gamma_E \e) \mu^{2\e} \, C^{-1-\e}]
\ee
where $C$ and $C^{-1-\e}$ are given by
\bea
\label{cs}
C &=& m_c^2 - 2 x y (qr) -  x (1-x) r^2 -i \delta \nonumber \\
C^{-1-\e} &=& - \exp(i \pi \e) \,  [x(1-x)]^{-1-\e} \,
\left[ r^2 + \frac{2 y (qr)}{1-x} - \frac{m_c^2}{x(1-x)} + i \delta
\right]^{-1-\e} \quad .
\eea
The range of integration in $(x,y)$ is restricted to the
simplex $S$, i.e.,  $0 \le y \le (1-x)$ and $0 \le x \le 1$.

Due to Ward identities,
not all the $\Delta i$ are independent.
The identities given in ref. \cite{Simma_Wyler}
in the context of the full theory simplify in our case
as follows:
\be
\label{ward}
q^\a \, J_{\a\b} = 0 \quad ; \quad
r^\b \, J_{\a\b} = 0 \quad .
\ee
They allow to express $\Delta i_5$ and
$\Delta i_6$ in terms of the other $\Delta i$
which have a more compact form.
These relations read
\bea
\label{substitut}
\Delta i_5 &=& \Delta i_{23}  \quad , \nonumber \\
\Delta i_6 &=& \frac{r^2}{(qr)} \, \Delta i_{25}
 + \Delta i_{26} \quad.
\eea
Of course, eq. (\ref{substitut})
can be checked explicitly for all values of $\e$,
using partial integration and certain
symmetry properties of the integrand.

We are now ready to evaluate the two-loop diagrams.
As both  $I_\b$ and $J_{\a\b}$
are transverse with respect to
the gluon, the gauge of the gluon propagator
is irrelevant.
Also, due to the absence of extra singularities in the limit of
vanishing strange quark mass, we set $m_s=0$ from
the very beginning (the question of charm quark mass
``singularities'' will be discussed later).

As an example,
we present the calculation of the two-loop
diagram in Fig. 1c in some detail.
Using $I_\b$
in eq. (\ref{build1}), the matrix element reads
\bea
\label{m1cc}
M(1c) &=&
\frac{i}{4 \pi^2} \, e Q_d g_s^2 C_F \,
\Gamma(\e) \exp(2 \g_E \e) \mu^{4 \e}
(1-\e) \exp(i \p \e) (4\p)^{-\e}\,  \int \,
\frac{d^dr}{(2\pi)^d}
\nonumber \\ &&
\bar{u}(p') \g^\b \frac{\pssl - \rsl }{(p'-r)^2} \, \epsl \,
\frac{\pbsl -\rsl }{(p-r)^2} \left(
r_\b \rsl - r^2 \g_\b \right) \, L \, u(p) \frac{1}{r^2}
\nonumber \\ &&
\int_0^1 \, dx \, \frac{[x(1-x)]^{1-\e}}{[r^2 - m_c^2/(x(1-x))
+ i \delta]^\e} \quad .
\eea
In eq. (\ref{m1cc}), $u(p')$ and $u(p)$ are the Dirac spinors
for the $s$ and the $b$ quarks, respectively, and
$C_F=4/3$.
In the next step, the four propagator factors in the denominator
are Feynman parametrized as
\be
\label{feynman}
\frac{1}{D_1 D_2 D_3 D_4^\e} = \frac{\G(3+\e)}{\G(\e)} \,
\int \frac{du dv dw dy \, y^{\e-1} \, \delta(1-u-v-w-y)}{
\left[D_1 u + D_2 v + D_3 w + D_4 y \right]^{3+\e}}
\quad ,
\ee
where
$D_1=(p'-r)^2$, $D_2=(p-r)^2$, $D_3=r^2$ and $D_4=r^2-m_c^2/(x(1-x))$.
Then the integral over the loop momentum $r$ is performed.
Making use of the $\d$ function in eq. (\ref{feynman}), the integral
over $w$ is easy. The remaining variables $u$, $v$ and
$y$ are transformed into new variables $u'$, $v'$ and $y'$,
all of them  varying in the interval [0,1]. The substitution reads
\be
\label{subst}
u \to v'(1-u') \quad , \quad
v \to v' u'\quad , \quad
y \to (1-v') y'  \quad .
\ee
Taking into account the corresponding jacobian
and omitting the primes($'$) of the integration variables
this leads to
\bea
\label{m1ca}
M(1c) &=&
\frac{1}{64 \pi^4} \, e Q_d g_s^2 C_F \,
\Gamma(2\e) \exp(2 \g_E \e) \mu^{4 \e} \,
\exp(2 i \pi \e) \,
(1-\e)  \, \int \, dx \,
du \, dv \, dy \times \nonumber \\
&& [x(1-x)]^{1-\e} \, y^{\e-1} \, (1-v)^\e \, v \,
\nonumber \\
&& \bar{u}(p') \,
\left[ P_1 \frac{\hat C}{\hat C^{2\e}} + P_2 \frac{1}{\hat C^{2\e}}
+ P_3 \frac{1}{\hat C^{1+2\e}} \right] \, u(p) \quad ,
\eea
where $P_1$, $P_2$ and $P_3$ are matrices in Dirac space
depending on the
Feynman parameters $x$, $u$, $v$, $y$
in a polynomial way.
$\hat C$ is given by
\be
\label{chat}
\hat C = m_b^2 v (1-v) u - \frac{m_c^2}{x(1-x)} (1-v) y + i \delta
\quad .
\ee
In what follows, the ultraviolet $\e$ regulator remains a fixed,
small positive number.

The central point of our procedure
is to use now the Mellin-Barnes representation of the "propagator"
 $1/(k^2 -M^2)^\l$ \cite{Boos,Usyukina,Smirnov,Erdelyi}
which is given by
\be
\label{Mellin}
\frac{1}{(k^2 - M^2)^\l} = \frac{1}{(k^2)^\l} \,
\frac{1}{\G(\l)} \, \frac{1}{2 \pi i} \, \int_{\gamma} ds
(-M^2/k^2)^s \G(-s) \G(\l+s) \quad ,
\ee
where $\l>0$ and
$\gamma$ denotes the integration path which is parallel to the
imaginary axis (in the complex $s$-plane) hitting the real axis
somewhere between $-\l$ and $0$. In this formula,
the "momentum squared" $k^2$ is understood
to have a small positive imaginary part.
In ref. \cite{Boos,Smirnov}
exact solutions to Feynman integrals
containing massive propagators are obtained
by representing their denominators
according to the formula (\ref{Mellin}) with subsequent
calculation of the corresponding massless integrals.

In our approach,
we use  formula (\ref{Mellin}) in order to simplify the
remaining Feynman parameter integrals in eq.
(\ref{m1ca}).  We represent
 the factors
$1/\hat{C}^{2\e}$ and $1/\hat{C}^{1+2\e}$ in eq.
(\ref{m1ca}) as Mellin-Barnes integrals using the identifications
\be
\label{ident}
k^2 \leftrightarrow m_b^2 v (1-v) u \quad ; \quad
M^2 \leftrightarrow \frac{m_c^2}{x(1-x)} (1-v) y \quad .
\ee
By interchanging the order of integration,
we first carry out the integrals over the Feynman parameters
for any given fixed value of $s$ on the integration path $\g$.
These integrals are basically the same as for the massless
case $m_c=0$ (in eqs. (\ref{m1ca}) and (\ref{chat})) up to
a factor (in the integrand) of
\be
\label{factor}
\left[ \frac{y}{u \, v \, x(1-x)} \right]^s
\, \exp(i \p s) \, \left( \frac{m_c^2}{m_b^2} \right)^s
\quad.
\ee
 Note that the polynomials $P_1$, $P_2$ and $P_3$
have such a form that the Feynman parameter integrals exist
in the limit $m_c \to 0$.
If the integration path $\g$ is chosen close enough to the
imaginary axis, the factor in eq. (\ref{factor}) does not change
the convergence properties of the integrals, i.e., the Feynman
parameter integrals exist for all values of $s$ lying on $\g$.
It is easy to see that the only
integrals involved are of the type
\be
\label{inttype}
\int_0^1 dw w^p \quad \mbox{or} \quad
\int_0^1 dw w^p (1-w)^q = \beta(p+1,q+1) = \frac{\G(p+1) \G(q+1)}{
\G(p+q+2)} \quad .
\ee
For the $s$ integration we use the residue
theorem after
closing the integration path
in the right $s$-halfplane.
One has to show that the integral over the half-circle
vanishes if its radius goes to $\infty$.
As we explicitly checked,
this is indeed the
case for $(m_c^2/m_b^2) < 1/4$, which is certainly satisfied
in our application.
The poles which lie
inside the
integration contour are
located at
\bea
\label{poles}
s &=& 0, 1, 2, 3, 4, ........ \nonumber \\
s &=& 1-\e, 2-\e, 3-\e, 4-\e, ....... \nonumber \\
s &=& 1-2\e, 2-2\e, 3-2\e, 4-2\e, ....... \quad .
\eea

The other two-loop
diagrams are evaluated similarly.
The non-trivial Feynman integrals can always be reduced to
those given in eq. (\ref{inttype})
after some  suitable substitutions.
The only change is that there are
poles in addition to those given in eq. (\ref{poles})
in those diagrams where the gluon
hits the $b$-quark line; they are located at
\be
\label{polesnew}
s=1/2-2\e, s=3/2-2\e, s=5/2-2\e, s=7/2-2\e,.... \quad .
\ee

The sum over the residues naturally leads to an
expansion in $z=(m_c^2/m_b^2)$ through the factor
$(m_c^2/m_b^2)^s$
in eq. (\ref{factor}).
This expansion, however, is not a Taylor series
because it also involves logarithms of $z$, which are generated
by the expansion in $\e$.
A generic diagram
which we denote by $G$ has then the form
\be
\label{generic}
G = c_0 + \sum_{n,m} c_{nm} z^n
\log^m z \quad ,  \quad z = \frac{m_c^2}{m_b^2} \quad ,
\ee
where the coefficients $c_0$ and $c_{n m}$ are independent
of $z$.
The power $n$ in eq. (\ref{generic})
is in general a natural multiple of $1/2$
and $m$ is a natural number including 0. In the explicit
calculation, the lowest $n$ turns out to be $n=1$.
This implies the important fact
 that the limit $m_c \to 0$ exists;
thus, there cannot be large logarithms (from a
small up-quark mass) in these diagrams.

From the structure of the poles one can see that the power
$m$ of the logarithm is bounded by  $4$
independent of the value of $n$. To illustrate this, take $n=100$
as an example. There are 3 poles located near n=100, viz.,
at $s=100, s=100 -\e, s=100-2\e$, respectively (see eq. (\ref{poles})).
Taking the residue at one of them, yields a term proportional to
$1/\e^2$ coming from the remaining two poles. In addition
there can be an explicit $1/\e^2$ term from the integration
of the two loop momenta. Therefore the most singular term can be
$1/\e^4$. Multiplying this with $z^s$ in eq. (\ref{factor})
leads to $z^{100} \log^m z$ where $m$ can be 4 at most.

We  have retained all terms up to $n=3$.
Comparing the $n=3$ numerical result with the one obtained
by truncating at $n=2$ leads to a difference of about $1\%$
only.

We have made further checks of our procedure.
For example we have calculated
diagram 1b directly. Expanding the result,
we reproduce
the expressions  obtained by applying the Mellin-Barnes integral at the
Feynman parameter level as described above. A similar
exercise for the imaginary part of diagram
1c shows that the
exact and the expanded result (up to $z^3$ terms)
in these examples agree at the $1\%$ level.
In addition, we checked that the imaginary part of the sum
of all diagrams coincides numerically with the results of
Soares \cite{Soares1,Soares2} (note, however that in
the physical region only the diagrams in
Fig. \ref{fig:1} and Fig. \ref{fig:3} contribute to the 
imaginary part).
In ref. \cite{Soares2}, 
Soares applied dispersion techniques 
to calculate the real part. 
However, using the imaginary part in the physical 
region, only the real part of the 
diagrams in Fig. \ref{fig:1} and
Fig. \ref{fig:3} is obtained. We have 
checked that our 
numbers for these two sets of diagrams indeed coincide with the results
of Soares. However, the contribution of the diagrams
in Fig. \ref{fig:2} and Fig. \ref{fig:4} is missing in \cite{Soares2}
because the additional 
unphysical cuts were not taken into account. 
We also note that a separate consideration of the
subtraction terms would be required
to obtain the correct $\mu$ dependence.

We mention that the Dirac algebra has been done with
the algebraic program REDUCE
\footnote{Some checks have been done with TRACER \cite{Tracer}}
\cite{Reduce}.
The Feynman parameter integrals and the determination of the
residues have been done with the symbolic program MAPLE \cite{Maple}.

We now give the results for the diagrams shown in Figs. 1 to 4.
As mentioned already, the individual diagrams in each figure
are not gauge invariant but only their sum.
Note, that the leading ultraviolet
singularity in the individual two-loop diagrams is in
general of order
$1/\e^2$. In the gauge invariant sums $M_2(i)$
the $1/\e^2$ cancel and we are left with $1/\e$-poles only.
The results read (using $z=(m_c/m_b)^2$ and $L=\log z$):
\bea
\label{m21res}
M_2(1) &=& \left\{ \frac{1}{36 \e} \,
\left( \frac{m_b}{\mu} \right)^{-4\e} \,
+ \frac{1}{216} \left[ 37 - (540 + 216 L) z \right. \right. \nonumber \\
&& \hspace{0.3cm}
+ (216 \p^2 - 540 +216 L - 216 L^2) z^2  \nonumber \\
&& \hspace{0.3cm} \left.
- (144 \p^2 +136 +240 L -144 L^2) z^3 \right] \nonumber \\
&& \hspace{0.3cm} \left.
+ \frac{i \pi}{18} \left[ 1 - 18 z +
(18-36 L) z^2 + (24 L -20) z^3 \right] \right\}
\nonumber \\
&& \hspace{0.3cm} \times
\frac{\a_s}{\p} \, C_F Q_d \bra s \g |O_7| b \ket _{tree}
\eea
\bea
\label{m22res}
M_2(2) &=&  \left\{ - \frac{5}{36 \e} \,
\left( \frac{m_b}{\mu} \right)^{-4\e} \,
+ \frac{1}{216} \left[ 13 + (36 \p^2 -108) z \right. \right. \nonumber \\
&& \hspace{0.3cm}
-144 \p^2 z^{3/2} + (648 - 648 L + 108 L^2) z^2  \nonumber \\
&& \hspace{0.3cm} \left. \left.
+ (120 \p^2 -314 +12 L +288 L^2) z^3 \right] \right\}
\nonumber \\
&& \hspace{0.3cm} \times
\frac{\a_s}{\p} \, C_F Q_d \bra s \g |O_7| b \ket _{tree}
\eea
\bea
\label{m23res}
M_2(3) &=&  \left\{ - \frac{1}{8 \e} \,
\left( \frac{m_b}{\mu} \right)^{-4\e} \,
+ \frac{1}{48} \left[ -45  \right. \right. \nonumber \\
&& \hspace{0.3cm}
+(72 -12 \pi^2 -96 \zeta (3) + (96 -24 \pi^2) L +12 L^2 + 8L^3) z
\nonumber \\
&& \hspace{0.3cm}
+(60 + 24 \pi^2 -96 \zeta(3) + (24 -24 \pi^2) L -24 L^2 + 8L^3) z^2
\nonumber \\ && \hspace{0.3cm} \left.
-(68-48L) z^3
\right] \nonumber \\
&& \hspace{0.3cm}
+ \frac{i \pi}{12} \left[ -3 +
(24 - 2 \p^2 + 6 L + 6 L^2) z  \right. \nonumber \\ && \hspace{0.3cm}
\left. \left. +
(6-2\pi^2-12 L +6 L^2) z^2 + 12 z^3 \right] \right\}
\nonumber \\
&& \hspace{0.3cm} \times
\frac{\a_s}{\p} \, C_F Q_u \bra s \g |O_7| b \ket _{tree}
\eea
\bea
\label{m24res}
M_2(4) &=&  \left\{ - \frac{1}{4 \e} \,
\left( \frac{m_b}{\mu} \right)^{-4\e} \,
- \frac{1}{24} \left[ 21   \right. \right. \nonumber \\
&& \hspace{0.3cm}
+(-24+2 \pi^2 +24 \zeta (3) - (12 -6 \pi^2) L + 2L^3) z
\nonumber \\
&& \hspace{0.3cm}
+(-12- 4 \pi^2 -48 \zeta(3) +  12 L -6 L^2 + 2L^3) z^2
\nonumber \\ && \hspace{0.3cm} \left. \left.
+(-6+6\pi^2-24L +18 L^2) z^3
\right] \right\}
\nonumber \\
&& \hspace{0.3cm} \times
\frac{\a_s}{\p} \, C_F Q_u \bra s \g |O_7| b \ket _{tree}
\eea
In these expressions, the symbol $\zeta$ denotes the Riemann
Zeta function, with $\zeta(3) \approx 1.2021$;
$Q_u=2/3$ and $Q_d=-1/3$ are the charge factors
for up- and down-type quarks, respectively.
The matrix element $\bra s \g |O_7| b \ket _{tree}$ is the
$O(\a_s^0)$ tree level
matrix element of the operator $O_7$;
its explicit form is
\be
\label{o7tree}
\bra s \g |O_7| b \ket _{tree} = m_b \, \frac{e}{8\p^2} \,
\bar{u}(p') \, \epsl \qsl R \, u(p) \quad .
\ee
In formula (\ref{o7tree}) $m_b$ should be identified with
the running mass $m_b(\mu)$ in principle
(see eq. (\ref{operators})).
However, as the corrections to $O_2$ are explicitly proportional
to $\a_s$, $m_b$ can be identified with the pole mass as well
(apart from $O(\a_s^2)$ corrections which we systematically neglect).

\subsection{Counterterms}
The operators mix under renormalization and thus the
counterterm
contributions must be taken into account. As we are 
interested in this
section in contributions to $b \to s \g$ which are proportional to
$C_2$, we have to include, in addition to the two-loop matrix elements
of $C_2 O_2$, also the one-loop matrix elements of the four Fermi
operators
$C_2 \delta Z_{2j} O_j$ ($j=1,...,6$) and the tree level contribution of
the magnetic operator $C_2 \delta Z_{27} O_7$.
In the NDR scheme  the
only non-vanishing contributions
to $b \to s \g$ come from  $j=5,6,7$. (For $j=5,6$ the
contribution comes from the diagram
in which the internal $b$-quark emits the photon).
The operator renormalization
constants $Z_{ij}$
are listed in the literature \cite{Ciuchini}
in the context of the leading
order anomalous dimension matrix.
The entries needed in our calculation
are
\bea
\label{zfactors}
\delta Z_{25} &=&  -\frac{\a_s}{48 \pi \e} \, C_F \quad , \quad
\delta Z_{26} =   \frac{\a_s}{16 \pi \e} \, C_F \quad ,
\nonumber \\
\delta Z_{27} &=&  \frac{\a_s}{16 \pi \e} \,
(6 Q_u - \frac{8}{9} Q_d) \, C_F \quad .
\eea
Defining
\be
M_{2j} = \bra s \g |\delta Z_{2j} O_j|b \ket \quad ,
\ee
we find the following contributions to the matrix elements
\bea
\label{counter}
M_{25} &=& -\frac{\a_s}{48 \pi} \, Q_d C_F \,
  \frac{1}{\e} \left(
\frac{m_b}{\mu} \right)^{-2\e}
\, \bra s \g |O_7| b \ket _{tree} \nonumber \\
M_{26} &=& \frac{3 \a_s}{16 \pi} \, Q_d C_F \,
  \frac{1}{\e} \left(
\frac{m_b}{\mu} \right)^{-2\e}
\, \bra s \g |O_7| b \ket _{tree} \nonumber \\
M_{27} &=& \frac{\a_s}{ \pi} \,
\left( \frac{3 Q_u C_F}{8} - \frac{Q_d C_F}{18} \right)
\, \frac{1}{\e} \, \bra s \g |O_7| b \ket _{tree}
\eea

We note that there is no one-loop contribution to 
the matrix element for $b \to s \gamma$ from the counterterm
proportional to $C_2 \, \frac{1}{\e} \, O_{12}^{ev}$, where
the evanescent operator $O_{12}^{ev}$ 
(see e.g. the last ref. in \cite{Ciuchini}) reads
\be
\label{evndr}
O_{12}^{ev} = \frac{1}{6}\, O_2 \left( \g_\mu \to
\g_{[\mu} \g_\nu \g_{\rho ]} \right) - O_2 \quad .
\ee
\subsection{Renormalized contribution proportional to
$C_2$}
\begin{figure}[htb]
\vspace{0.10in}
\centerline{
\epsfig{file=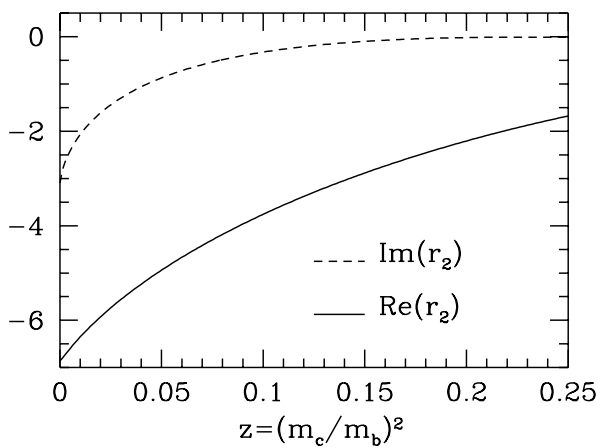,height=2.5in,angle=0,clip=}
}
\vspace{0.08in}
\caption[]{
Real and imginary part of $r_2$ in the NDR scheme
(from eqs.
(\ref{rer2ndr}) and (\ref{imr2ndr})).
\label{r2figure}}
\end{figure}

Adding the two-loop diagrams from section 2.1 (eqs.
(\ref{m21res}), (\ref{m22res}), (\ref{m23res}) and (\ref{m24res}))
and the counterterms from section 2.2
(eq. (\ref{counter})) we find the renormalized contribution
$M_2$
\be
\label{m2formal}
M_2 = M_2(1) + M_2(2) + M_2(3) + M_2(4) + M_{25}
+ M_{26} + M_{27} \quad .
\ee
Of course, the ultraviolet singularities cancel
in $M_2$. Inserting $C_F=4/3$,
$Q_u=2/3$ and
$Q_d=-1/3$, we get the main result of this paper,
which in the NDR scheme reads:
\be
\label{m2lr}
M_2 = \bra s \g |O_7| b \ket _{tree} \, \frac{\a_s}{4 \p} \,
\left( \ell_2 \log \frac{m_b}{\mu}  + r_2 \right) \quad ,
\ee
with
\be
\label{l2}
\ell_2 = \frac{416}{81}
\ee
\bea
\label{rer2ndr}
\Re r_2 &=& \frac{2}{243} \, \left\{- 833 + 144 \pi^2 z^{3/2}
\right. \nonumber \\
&& \hspace{0.3cm}
+ \left[ 1728 -180 \pi^2 -1296 \zeta (3) + (1296-324 \pi^2) L +
108 L^2 + 36 L^3 \right] \, z \nonumber \\
&& \hspace{0.3cm}
+ \left[ 648 + 72 \pi^2 + (432 - 216 \pi^2) L + 36 L^3 \right] \, z^2
\nonumber \\
&& \hspace{0.3cm}        \left.                 +
\left[ -54 - 84 \pi^2 + 1092 L - 756 L^2 \right] \, z^3 \, \right\}
\eea
\bea
\label{imr2ndr}
\Im r_2 &=& \frac{16 \p}{81} \, \left\{- 5
+ \left[ 45-3 \pi^2 + 9 L +
9 L^2 \right] \, z
+ \left[ -3 \pi^2 + 9 L^2 \right] \, z^2 +
\left[ 28 - 12 L  \right] \, z^3 \, \right\}
\eea
Here, $\Re r_2$ and $\Im r_2$ denote the real and the imaginary part
of $r_2$, respectively. The quantity $z$ is defined as $z=(m_c^2/m_b^2)$
and $L=\log(z)$.
In Fig. \ref{r2figure}
we show the real and the imaginary part of
$r_2$.
For  $z \ge 1/4$ the imaginary part must
vanish exactly; indeed  we see
from Fig. \ref{r2figure} that the imaginary part
based on the expansion retaining terms up to
$z^3$
indeed
vanishes at $z=1/4$ to high accuracy.

\section{$O(\a_s)$ corrections to $O_7$}
\begin{figure}[htb]
\vspace{0.10in}
\centerline{
\epsfig{file=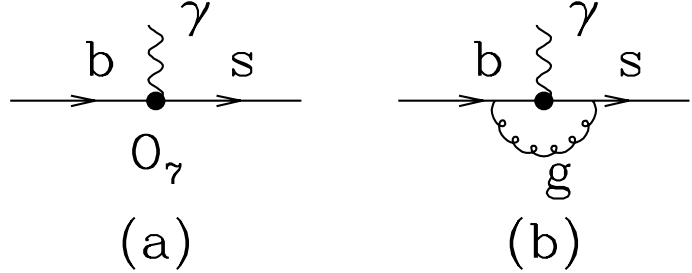,height=1.5in,angle=0,clip=}
}
\vspace{0.08in}
\caption[]{
Virtual corrections to $O_7$
\label{fig:7}}
\end{figure}

\setcounter{equation}{0}
The virtual corrections associated with the
operator $O_7$ as shown in Fig. \ref{fig:7}b (together with
the selfenergy diagrams and the counterterms)
have been taken into account
in the work of Ali and Greub, see e.g. \cite{agalt,aglett,aglong},
where $m_s \neq 0$ was retained.
Since we neglect $m_s$ in this work, we are interested only in the
limit $m_s \to 0$.
Because of the mass
singularities in the virtual corrections (which will be cancelled
when also taking into account Bremsstrahlung corrections),
we only keep $m_s$ as a regulator.

Including the lowest order contribution, the
result then becomes (using $\rho=(m_s/m_b)^2$) in the NDR scheme
\be
\label{o7virt}
\bra s \g|O_7| b \ket _{virt} = \bra s \g|O_7| b \ket _{tree}
\, \left[ 1 + \hat{K}_g   \right] \quad ,
\ee
\be
\label{krenorm}
\hat{K}_g = \frac{\a_s}{6 \p} \,
\left( \frac{4 \p \mu^2}{m_b^2} \right)^{\e_{IR}} \,
\G(1+\e_{IR}) \, \left\{ \log^2 \rho - \frac{2}{\e_{IR}}
\log \rho - \log \rho -
\frac{4}{\e_{IR}} - 8 + 4 \log \frac{m_b}{\mu}.
\right\}
\ee
Note, that
eq. (\ref{o7virt}) contains all the counterterm contributions.
The $1/\e_{IR}$-poles in this equation are therefore of infrared
origin as indicated by the notation. The last term in the curly
bracket in eq. (\ref{krenorm}) represents a $\mu$ dependence of
ultraviolet origin. The additional $\mu$ dependence, which is generated
when expanding in $\e_{IR}$ is cancelled at the level of the decay
width
together with the
$1/\e_{IR}$-poles when adding the Bremsstrahlung correction
due to the square of the diagrams associated
with the operator $O_7$.
As all intermediate formulae are given in the literature,
we only give the final result for the $\a_s$ corrections
(virtual+Bremsstrahlung)
to the decay width. Denoting this contribution by
 $\G_{77}$ we get
in the limit
$m_s=0$
\be
\label{Gamma7}
\G_{77} = \G_{77}^0 \,
\left[ 1 + \frac{\a_s}{3 \p} \left(
\frac{16}{3} - \frac{4\p^2}{3} + 4 \log \frac{m_b}{\mu} \right)
\right]  \quad ,
\ee
where  the lowest order contribution $\G_{77}^0$ reads
\be
\label{Gamma70}
\G_{77}^0 = \frac{m_b^2(\mu) m_b^3}{32 \p^4} \,
|G_F \l_t C_7^{eff}|^2 \,
\a_{em} \quad .
\ee

For later convenience,
we can formally rewrite $\bra s \g|O_7| b \ket _{virt} $
in eq. (\ref{o7virt}) in such a way that its square reproduces
the result in eq. (\ref{Gamma7}). This modified matrix element,
denoted by $\bra s \g|O_7| b \ket _{rad}$ then reads
\be
\label{m7lr}
\bra s \g|O_7| b \ket _{rad} = \bra s \g|O_7| b \ket _{tree}
\, \left( 1+ \frac{\a_s}{4\p} \left( \ell_7 \, \log \frac{m_b}{\mu}
+r_7
\right) \right)
\ee
with
\be
\label{l7r7}
\ell_7 = \frac{8}{3}  \quad , \quad
r_7 = \frac{8}{9} \, (4 - \pi^2) \quad .
\ee

\section{Virtual corrections to  $O_8$}
\begin{figure}[htb]
\vspace{0.10in}
\centerline{
\epsfig{file=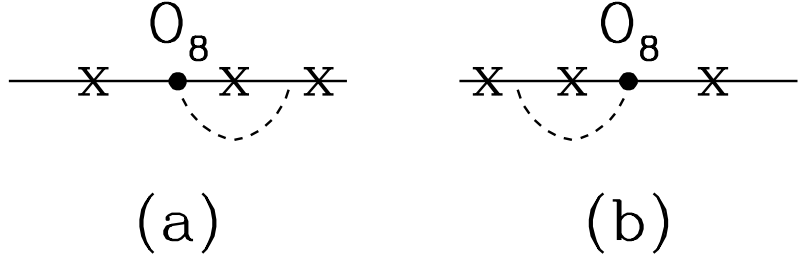,height=1.5in,angle=0,clip=}
}
\vspace{0.08in}
\caption[]{Contributions of $O_8$ to $b \to s \g$. The cross (x)
denotes the possible place where the photon is emitted. Figs.
(a) and
(b) are separately gauge invariant.
\label{fig:8}}
\end{figure}

\setcounter{equation}{0}
Finally, we consider the contributions to $b \to s \g$
generated by the operator $O_8$, i. e.,
\be
\label{m8def}
M_8 = \bra s \g|O_8|b \ket \quad.
\ee
The corresponding Feynman diagrams
are shown in Fig. \ref{fig:8}.
The sum of the three diagrams in Fig. \ref{fig:8}a yields
\be
\label{ress123}
M_8(9a) = \frac{Q_d \, C_F}{2} \, \frac{\a_s}{\p} \,
\left[ -\frac{1}{\e} -2 +2 \log(m_b/\mu) - i \p \right] \,
\bra s \g|O_7|b \ket _{tree}
\quad ,
\ee
while the diagrams in Fig. \ref{fig:8}b gives
\be
\label{ress456}
M_8(9b) = \frac{Q_d \, C_F}{12} \, \frac{\a_s}{\p} \,
\left[ -\frac{6}{\e} -21 +2\p^2 +12 \log(m_b/\mu) \right]  \,
\bra s \g|O_7|b \ket _{tree}
\quad .
\ee
Thus the sum of all the six diagrams is
\be
\label{ress123456}
M_8(9) = \frac{Q_d \, C_F}{12} \, \frac{\a_s}{\p} \,
\left[ -\frac{12}{\e} -33 +2\p^2 +24 \log(m_b/\mu) -6 i \p
\right]  \,
\bra s \g|O_7|b \ket _{tree}
\quad .
\ee
There is also a contribution from a counterterm; it
reads
\be
\label{m8counter}
M_{87} = \delta Z_{87} \,
\bra s \g|O_7|b \ket _{tree}   \quad ,
\ee
The renormalization constant
\be
\delta Z_{87} =  \frac{\a_s}{\p} \, C_F Q_d \,\frac{1}{\e}
\ee
has been calculated in the literature \cite{Ciuchini}.
The sum of all contributions leads to the
renormalized result $M_8$:
\be
\label{m8lr}
M_8 = \bra s \g |O_7| b \ket _{tree} \, \frac{\a_s}{4 \p} \,
\left( \ell_8 \log \frac{m_b}{\mu}  + r_8 \right) \quad ,
\ee
with
\be
\label{l8r8}
\ell_8 = - \frac{32}{9} \quad , \quad
r_8 = - \frac{4}{27} \,
\left( -33 +2\p^2  -6 i \p
\right)
\quad .
\ee

\section{Results and conclusions}
\setcounter{equation}{0}
We have calculated the virtual
corrections to $b \to s \gamma$ coming from the operators
$O_2$, $O_7$ and $O_8$. The contributions from the other
four-Fermi operators 
in eq. (\ref{operators}) which are given by the analogous diagrams
as shown in Figs. (\ref{fig:1}) -- (\ref{fig:4}) were
neglected, because they either
vanish ($O_1$) or have Wilson coefficients 
which are about fifty times
smaller than that of $O_2$ while their matrix elements
can be enhanced at most by color factors.
However, we did include the non-vanishing
diagrams of  $O_5$ and $O_6$
where the gluon connects the external
quark lines and the photon is radiated from the charm quark
because these corrections are 
automatically considered
when $C_7^{eff}$ (defined in eq.(\ref{C78eff}))
is used instead of $C_7$.
As discussed in section three, some of the Bremsstrahlung
corrections to  the operator $O_7$ have been transferred into
the matrix element for $b \to s \gamma$ in order 
to present results which are free from 
infrared and collinear singularities.

The sum of the various contributions derived 
in the previous sections yields the amplitude $A(b \to s \g)$
for $b \to s \gamma$. The result can be presented in
a convenient way, following the treatment of
Buras et al. \cite{Buras94}, where the general structure
of the next-to-leading order result is discussed in detail.
We write
\be
\label{amplitudevirt}
A(b \to s \g) = -\frac{4 G_F \l_t}{\sqrt{2}} \, \hat{D} \,
\bra s \g|O_7(\mu)|b \ket _{tree}
\ee
with
$\hat{D}$
\be
\label{dhat}
\hat{D} = C_7^{eff}(\mu) + \frac{\a_s(\mu)}{4\p} \sum_{i}
\left(
C_i^{(0)eff}(\mu) \ell_i \log \frac{m_b}{\mu} +
C_i^{(0)eff} r_i
\right)         \quad ,
\ee
and where the quantities $\ell_i$ and $r_i$ are given
for $i=2,7,8$ in sections 2, 3 and 4, respectively.
For the full next-to-leading logarithmic result one would need  
the first term on the rhs of eq. (\ref{dhat}) ,$C_7^{eff}(\mu)$,
at next-to-leading logarithmic  
precision.  In contrast, 
it is consistent 
to use the leading logarithmic values
for the other Wilson coefficients in eq. (\ref{dhat}).
As the next-to-leading coefficient $C_7^{eff}$ is not known
yet, we replace it by its
leading logarithmic value $C_7^{(0)eff}$ 
in the numerical investigations.
The notation $\bra s \g|O_7(\mu)|b \ket _{tree}$ in
eq. (\ref{amplitudevirt}) indicates that the explicit 
factor $m_b$
in the operator $O_7$ is the running mass taken
at the scale $\mu$.

As the relevant scale for a $b$ quark decay is expected to
be $\mu \sim m_b$, we expand the matrix elements of the
operators around
$\mu=m_b$ up to order $O(\a_s)$.
Thus we arrive at 
\be
\label{amplitudevirtuell}
A(b \to s \g) = -\frac{4 G_F \l_t}{\sqrt{2}} \, D \,
\bra s \g|O_7(m_b)|b \ket _{tree}
\ee
with
$D$
\be
\label{d}
D = C_7^{eff}(\mu) + \frac{\a_s(m_b)}{4\p} \sum_{i} \left(
C_i^{(0)eff}(\mu) \gamma_{i7}^{(0)eff} \log \frac{m_b}{\mu} +
C_i^{(0)eff} r_i
\right)         \quad ,
\ee
with the quantities $\gamma_{i7}^{(0)eff}$
\be
\gamma_{i7}^{(0)eff} = \ell_i + 8 \delta_{i7}
\ee
being just the entries of the (effective) leading order anomalous
dimension matrix \cite{Buras94}.
As also pointed out in this reference,
the explicit logarithms of the form
$\a_s(m_b) \log(m_b/\mu)$
in eq. (\ref{d})
should be cancelled by the $\mu$-dependence of $C_7^{(0)eff}(\mu)$.
This is the crucial point why the scale dependence
is reduced significantly as we will see later. 
\footnote{As we neglect the virtual correction of $O_3$-$O_6$,
there is a small left-over
$\mu$ dependence, of course.}

From $A(b \to s \g)$ in eq.
(\ref{amplitudevirtuell}) we obtain
the decay width $\G^{virt}$ to be
\be 
\label{widthvirt}
\G^{virt} = \frac{m_{b,pole}^5 \, G_F^2 \l_t^2 \a_{em}}{32 \p^4}
\, F \, |D|^2 \quad ,
\ee
where we discard term of $O(\a_s^2)$
in $|D|^2$.
The factor $F$ in eq. (\ref{widthvirt}) is
\be
F = \left( \frac{m_b(\mu=m_b)}{m_{b,pole}} \right)^2 =
1- \frac{8}{3} \,  \frac{\a_s(m_b)}{\p} \quad .
\ee

To obtain the inclusive rate for $B \to X_s \g$
consistently at the next-to-leading order level, 
we have to take into account all the Bremsstrahlung contributions.
They have been calculated by
Ali and Greub for the operators
$O_2$ and $O_7$ some time ago \cite{agalt}, while the complete set has been
worked out only recently
\cite{aglett,aglong,Pott}.
Here, we neglect the small 
contribution
of the operators $O_3$ -- $O_6$ as we did for the virtual corrections,
i.e., we only consider $O_2$, $O_7$ and $O_8$.
The  corresponding Bremsstrahlung formulae are collected 
in Appendix B.

In order to arrive at  the  branching ratio $\mbox{BR}(b \to s \g (g))$,
we divide, as usual, the decay width $\G(B \to X_s \g) = 
\G^{virt} + \G^{brems}$
by the theoretical expression for the
semileptonic decay width $\G_{sl}$ and
multiply this ratio with the measured semileptonic branching ratio
$\mbox{BR}_{sl} = (10.4 \pm 0.4)\%$ \cite{Gibbons}, i.e.,
\be
\label{brformel}
\mbox{BR}(b \to s \g (g)) = \frac{\G}{\G_{sl}} \,
\mbox{BR}_{sl} \quad .
\ee
\begin{figure}[htb]
\vspace{0.10in}
\centerline{
\epsfig{file=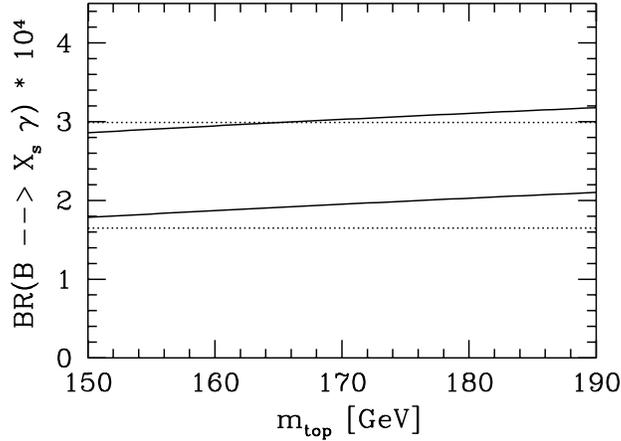,height=2.5in,angle=0,clip=}
}
\vspace{0.08in}
\caption[]{Branching ratio for $b \to s \g$
based on the leading logarithmic formula in eq.
(\ref{leadinglog}).
The upper (lower) solid curve
is for $\mu=m_b/2$ ($\mu = 2 m_b$). The dotted curves show the
CLEO bounds \cite{CLEOrare2}.
\label{fig:mtspecllog}}
\end{figure}
\begin{figure}[htb]
\vspace{0.10in}
\centerline{
\epsfig{file=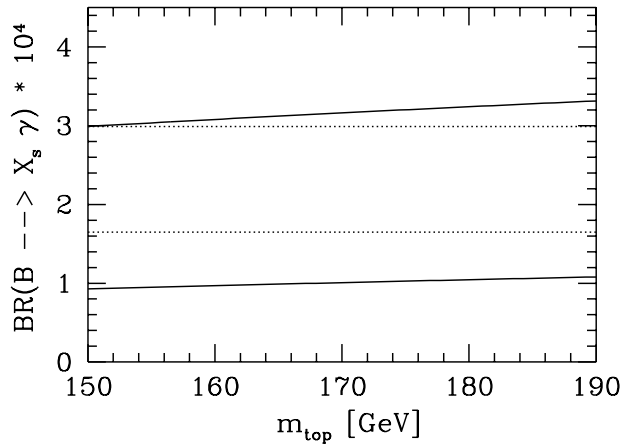,height=2.5in,angle=0,clip=}
}
\vspace{0.08in}
\caption[]{Branching ratio for $b \to s \g (g)$ neglecting
the virtual corrections of $O_2$ and $O_8$
calculated in the present
paper.
The upper (lower) solid curve
is for $\mu=m_b/2$ ($\mu = 2 m_b$).
The dotted curves show the CLEO bounds \cite{CLEOrare2}.
\label{fig:mtspecali}}
\end{figure}
\begin{figure}[htb]
\vspace{0.10in}
\centerline{
\epsfig{file=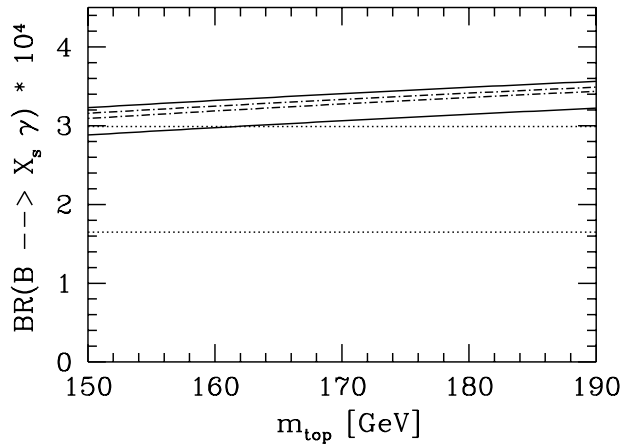,height=2.5in,angle=0,clip=}
}
\vspace{0.08in}
\caption[]{Branching ratio for $b \to s \g (g)$
based on the complete 
formulae presented in this section. The upper (lower) solid curve
is for $\mu=m_b/2$ ($\mu = 2 m_b$). The dotted curves show the CLEO
bounds \cite{CLEOrare2}.
\label{fig:mtspecfinal}}
\end{figure}
The semileptonic decay width is
\be
\label{semileptonic}
\G_{sl} = \frac{G_F^2 \, m_{b,pole}^5 \, |V_{cb}|^2}{192 \p^3} \,
g(m_c/m_b) \, \left( 1 -
\frac{2 \a_s(m_b)}{3 \p} f(m_c/m_b) \right) \quad ,
\ee
where the phase space function $g(u)$ is defined as
\be
\label{gz}
g(u) = 1 - 8 u^2 + 8 u^6 - u^8 - 24 u^4 \log u \quad ,
\ee
and an approximate analytic form
for the radiative correction function
$f(u)$ has been found
in \cite{CCM} to be
\be
\label{fz}
f(u) = \left( \p^2 - \frac{31}{4} \right) \, (1-u)^2 + \frac{3}{2}
\quad .
\ee

In Figs. \ref{fig:mtspecllog}-\ref{fig:mtspecfinal} we compare the
available leading-log results and our new results
for the inclusive  branching ratio for $B \to X_s \g$
as a function of the top quark mass.
A rather crucial parameter is the ratio $m_c/m_b$;
it enters both,
$b \to s \g$ through the
virtual corrections of $O_2$ and
the semileptonic decay width through phase space.
It can be written as $m_c/m_b=1-(m_b-m_c)/m_b$. 
While the mass difference $m_b - m_c$
is determined quite precisely through the $1/m_Q$
expansion \cite{HQET2} or from the semileptonic
$b \to c$ spectrum; we use $m_b - m_c=3.40$ GeV \cite{Shifmanetal},
the $b-$quark mass is not precisely known.
Using for the $b$ quark pole
mass $m_{b,pole}=4.8 \pm 0.15$ GeV one arrives at  $m_c/m_b=0.29 \pm 0.02$.
In the plots the central values for $m_{b,pole}$ 
and $m_c/m_b$ have been used. Moreover, we put $V_{cb}=V_{ts}$ and $V_{tb}=1$
and also use the central value for the measured semileptonic branching
ratio $\mbox{BR}_{sl}=10.4\%$ in eq. (\ref{brformel}).
In all three plots the horizontal dotted
curves show the CLEO $1 \sigma$-limits for the branching ratio
$\mbox{BR}(B \to X_s \g)$ \cite{CLEOrare2}. 

In Fig. \ref{fig:mtspecllog} we show the  
leading logarithmic
result for the branching ratio of $b \to s \g$,
based on the formula
\be
\label{leadinglog}
\mbox{BR}(b \to s \g)^{leading} =
\frac{6 \a_{em}}{\p g(m_c/m_b)} \, |C_7^{(0)eff}|^2 \,
\mbox{BR}_{sl}         \quad .
\ee
Similarly, Fig. \ref{fig:mtspecali} exhibits 
the results also
taking into 
account the Bremsstrahlung corrections and the virtual
corrections to $O_7$ without  including the virtual corrections of $O_2$
and $O_8$. We can 
reproduce this result by
putting $\ell_2=r_2=\ell_8=r_8=0$ in our formulae.
As noticed in the literature  \cite{agalt,Pott},
the $\mu$ dependence in this case is even larger
than in the leading
logarithmic result shown in Fig. \ref{fig:mtspecllog}.
Also, the experimentally allowed region is shown.

The theoretical results shown in  Figs. \ref{fig:mtspecllog} 
and
\ref{fig:mtspecali} 
allowed for a reasonable
prediction of the branching ratio within a large error
which was essentially determined by the $\mu$ dependence.
As we see, they do agree well with experiment but within 
large uncertainties.
 
In Fig. \ref{fig:mtspecfinal}, finally, we
give the branching ratio 
based on  formula (\ref{brformel}) which
includes all virtual corrections calculated
in the present paper.  
Because all the logarithms
of the form $\a_s(m_b) \log(m_b/\mu)$ cancel as discussed
above, the $\mu$ dependence is significantly reduced in our improved 
calculation (Fig. \ref{fig:mtspecfinal}); the bands of
scale uncertainty are rather narrow. 
To illustrate the remaining renormalization scale dependence,
we present two 'scenarios'
which differ by higher order contributions. First, we take the 
explicit $\alpha_s$ factors in eqs. (\ref{d}) 
and (\ref{semileptonic}) at $\mu=m_b$ as indicated in these
formulae; varying $\mu$ between $(m_b/2)$ and $(2 \, m_b)$
in formula (\ref{d}) leads to the 
dash-dotted curves in Fig. \ref{fig:mtspecfinal}. 
Second, we evaluate the explicit $\alpha_s$ in eqs. (\ref{d})
and (\ref{semileptonic}) at the (variable) scale $\mu$.
Varying again the scale $\mu$ between $(m_b/2)$ and $(2\,m_b)$
yields the solid lines in Fig. \ref{fig:mtspecfinal}.
In both scenarios
the upper (lower) curve corresponds to $\mu=m_b/2$ ($\mu=2m_b$).
We mention that the $\mu$-band is larger in the second scenario
and it is therefore safer to use this band to 
obtain a feeling for the remaining scale uncertainties.

While this result shows that the theoretical accuracy can
be strongly improved by the next-to-leading calculations,
it would be premature to extract a prediction
for the branching ratio from Fig. \ref{fig:mtspecfinal}
(with obviously a small error) and claim, for instance, 
a discrepancy with experiment. This only will become
possible (with high theoretical precision)
if also  $C_7^{eff}$  is known to next-to-leading
logarithmic precision. This additional effect will , essentially, 
shift the narrow bands in Fig. \ref{fig:mtspecfinal}, without
broadening them substantially.
The drastic reduction of the theoretical
uncertainties shows  that  significant experimental 
improvements are 
necessary to extract the important information 
available in these decays.

\vspace{2cm}
{\bf Acknowledgements}
We thank A. Ali, M. Beneke, R. Blankenbecler,
S. Brodsky, J. Hewett, M. Lautenbacher, M. Peskin, 
J. Soares 
and M. Worah for discussions.
We are particularly indebted to M. Misiak and L. Reina
for many useful comments. One of us (C.G.) would like to thank the
Institute for Theoretical Physics in Z\"urich
for the kind hospitality.

\newpage
\centerline{\Large {\bf Appendix}}
\appendix
\section{$O_2$ contribution in the `t Hooft-Veltman scheme}
\setcounter{equation}{0}
In this appendix we present the results for the matrix elements
of the process $b \to s \gamma$
based on the operator $O_2$. In addition to the two-loop
diagrams shown in Figs. \ref{fig:1} -- \ref{fig:4}, we also
give all the counterterm contributions which multiply the
Wilson coefficient $C_2$. 
\subsection{The `t Hooft-Veltman Scheme}
In the `t Hooft-Veltman scheme \cite{HV}, the $d$ dimensions are 
split into 4 and $d-4$; the corresponding structures
are distinguished by no superscript, by a tilde  and by a hat,
respectively. There are Lorentz indices in $d$, 4
and $d-4$ dimensions and the corresponding metric
tensors $g_{\mu \nu}$, $\tilde{g}_{\mu \nu}$ and $\hat{g}_{\mu \nu}$.
While all the $\g$-matrices are taken in $d$ dimensions, their indices
are split in $4$ and $d-4$ components, according to the rules
\bea
\label{HVrules}
g_{\mu \nu} &=& \tilde{g}_{\mu \nu} + \hat{g}_{\mu \nu} \nonumber \\
\tilde{g}_{\mu \nu} \tilde{g}^{\mu \nu} &=& 4 \quad , \quad 
\hat{g}_{\mu \nu} \hat{g}^{\mu \nu} = d-4 \quad , \nonumber \\
\tilde{g}_{\mu \nu} \hat{g}^{\mu \nu} &=& 0 \quad .
\eea
The $\g$-matrices in 4 dimensions ($\tilde{\g}^\mu$) and $(d-4)$
dimensions
($\hat{\g}^\mu$) are defined by $\tilde{g}^{\mu \nu} \g_\nu$
and $\hat{g}^{\mu \nu} \g_\nu$, respectively. Assuming the usual
anticommutation relations of the $d$-dimensional Dirac matrices
in terms of the $d$-dimensional metric tensor $g_{\mu \nu}$,
one gets the
following rules for $\tilde{\g}^\mu$ and $\hat{\g}^\mu$.
\be
\label{anticomm}
\left\{ \tilde{\g}^\mu, \tilde{\g}^\nu \right\} = 2 \tilde{g}^{\mu \nu}
\quad , \quad
\left\{ \hat{\g}^\mu, \hat{\g}^\nu \right\} = 2 \hat{g}^{\mu \nu}
\quad , \quad
\left\{ \tilde{\g}^\mu, \hat{\g}^\nu \right\} = 0 \quad .
\ee
The commutation relations with $\g_5$ are postulated to be
\be
\left\{ \tilde{\g}^\mu, \g_5 \right\} = 0
\quad , \quad
\left[ \hat{\g}^\mu, \g_5 \right] = 0
\quad ,
\ee
which is equivalent to defining $\g_5$ by the product
$i \tilde{\g_0} \tilde{\g_1} \tilde{\g_2} \tilde{\g_3} $.
This is the only way known to treat $\g_5$ without running into
algebraic inconsistencies
\cite{Breitenlohner}.
Finally, we mention that the chiral vertices in $d$-dimensions
can be defined in different ways, all having the same formal limit
when $d \to 4$. For left- and right- handed currents we follow
the common practice and
use
\be
\label{currents}
\tilde{\g}^\mu L = R \g^\mu L \quad \mbox{and} \quad
\tilde{\g}^\mu R = L \g^\mu R \quad .
\ee
There are several possibilities
to define the operators $O_7$ and $O_8$ in $d$ dimensions
(with identical 4 dimensional limit);
for example the term $\sigma^{\mu \nu}$
in eq.
(\ref{operators}) can be defined to be
\be
\label{def1}
\sigma^{\mu \nu} = \frac{i}{2} [\g^\mu,\g^\nu] \quad ,
\ee
where $\g^\mu$ and $\g^\nu$ are the "$d$-dimensional"
matrices, or, alternatively,
\be
\label{def2}
\sigma^{\mu \nu} = \frac{i}{2} [\tilde{\g}^\mu,\tilde{\g}^\nu] \quad .
\ee
A difference in the
definition will result in a difference of the finite terms
in the one-loop matrix elements of these operators.

As the calculation of the one-loop matrix elements of
$O_7$ and $O_8$
is relatively
easy, once an exact definition of the operators has been
specified, we give in this appendix only the result for the two-loop
contribution of the operator $O_2$.

\subsection{One-loop building blocks }
The principal steps of the calculation in the HV scheme
of the $O_2$ contribution to the matrix element
for $b \to s \gamma$ are the same as in the NDR scheme.
Therefore, we again start with the one-loop building blocks.
For the building block $I_\beta$ in Fig. \ref{fig:5}
we get in the HV scheme
\bea
\label{build1hv}
I_\beta &=& - \frac{g_s}{4 \pi^2} \, \Gamma(\e) \,
\mu^{2 \e} \,
\exp(\gamma_E \e)
\,
\exp(i \pi \e) \,
\int_0^1 \, dx [x(1-x)]^{-\e} \,
\left[ r^2 - \frac{m_c^2}{x(1-x)} + i \delta \right]^{-\e}
\nonumber \\
&& \left[ x(1-x) \,
\left(r_\b r^\g \tilde{\g}_\g - r^2 \tilde{\g}_\b \right) +
\frac{m_c}{2} r^\g \left(
\hat{\g}_\b \tilde{\g}_\g+ \tilde{\g}_\b \hat{\g}_\g
\right)  \right]
\, L
\frac{\lambda}{2}
\quad ,
\eea
while the building block $J_{\a \b}$ in Fig. \ref{fig:6}
reads
\bea
\label{build2hv}
J_{\a \b}  &=& \frac{e g_s Q_u}{32 \pi^2 \, (1+\e)} \, R \,
\tilde{\g}_\m \,
\left[ E(\a,\b,r)  \, \Delta i_5 +
       E(\a,\b,q) \, \Delta i_6
       - E(\b,r,q) \, \frac{r_\a}{(qr)} \Delta i_{23}
       \nonumber \right. \\ && \left. \hspace{1.0cm}
       - E(\a,r,q) \, \frac{r_\b}{(qr)} \Delta i_{25}
       - E(\a,r,q) \, \frac{q_\b}{(qr)} \Delta i_{26}
       \right] \, \tilde{\g}^\m \, L \,
\frac{\lambda}{2} \nonumber \\ && \hspace{1.0cm}
- m_c \, \frac{e g_s Q_u}{8 \pi^2} \, \frac{1}{(qr)} \,
\, R \,  \tilde{\g}_\m \, \left\{
\rsl \qsl g_{\a\b} - \g_\b \qsl r_\a - \g_\a \rsl q_\b -\g_\a 
\g_\b \rsl\qsl \right. \nonumber \\
&& \hspace{1.0cm}
\left. + \g_\a \g_\b (qr) \right\} \, \tilde{\g}^\m \, L \,
\frac{\l}{2} \, \hat{\Delta}
\quad .
\eea
The quantities $E$ and 
$\Delta i$ are given in eq.
(\ref{epsilongeneralization})
and (\ref{deltai5})-(\ref{deltai25}), respectively.
As the $\Delta i$ are the same as in the NDR
scheme,
the Ward identities in
eq. (\ref{substitut}) still hold.
The additional term $\hat{\Delta}$ in eq. (\ref{build2hv})
reads
\be
\label{deltazusatz}
\hat{\Delta} = \int_{S} \, dx \, dy \, (qr) \,
C^{-1-\e} \, \e \, \Gamma(\e) \, \mu^{2\e} \, e^{\gamma_E \e}
\quad ,
\ee
 where
$C^{-1-\e}$ and the integration 
range $S$ are defined in eq. (\ref{cs}). 

We note that the terms linear in $m_c$ in eqs. (\ref{build1hv})
and (\ref{build2hv}) are absent in schemes where $\gamma_5$
has formally the same anticommuting properties as in the 4-dimensional
Dirac
algebra. Especially, they are absent in the NDR scheme.
However, working consistently in the HV scheme, we have to retain these
extra terms. 

In order to make the comparison with the NDR calculation easier,
we discard in a first step these extra $m_c$ terms and denote again
the corresponding results by $M_2(i)$ 
where $i=1,2,3,4$ numbers the figure in which the 
corresponding diagrams are
shown. In sections A.3, A.4 and A.5 
the  terms linear in $m_c$ are left out;
they are denoted by $M^c_2(i)$ and collected
in section A.6.
Those counterterm contributions involving evanescent
operator which generate only terms linear in $m_c$ are 
also discussed there.

\subsection{Regularized two-loop contribution of $O_2$}
We now give the results for the two-loop diagrams in Figs.
\ref{fig:1} -- \ref{fig:4} discarding the linear terms in $m_c$ 
in the one-loop building blocks
in eqs. (\ref{build1hv}) and (\ref{build2hv}). The results read,
using
$z=(m_c/m_b)^2$ and $L=\log z$,
\bea
\label{m21reshv}
M_2(1) &=& \left\{ \frac{1}{36 \e} \,
\left( \frac{m_b}{\mu} \right)^{-4\e} \,
+ \frac{1}{216} \left[ 40 - (540 + 216  L) z \right. \right. \nonumber \\
&& \hspace{0.3cm}
+ (216 \p^2 - 540 + 216 L -  216 L^2) z^2  \nonumber \\
&& \hspace{0.3cm} \left.
- ( 144 \p^2 + 136 + 240 L - 144 L^2) z^3 \right] \nonumber \\
&& \hspace{0.3cm} \left.
+ \frac{i \pi}{18} \left[ 1 - 18 z +
(18-36 L) z^2 + (24 L -20) z^3 \right] \right\}
\nonumber \\
&& \hspace{0.3cm} \times
\frac{\a_s}{\p} \, C_F Q_d \bra s \g |O_7| b \ket _{tree}
\eea
\bea
\label{m22reshv}
M_2(2) &=& \left\{ -\frac{1}{18 \e} \,
\left( \frac{m_b}{\mu} \right)^{-4\e} \,
+ \frac{1}{216} \left[-17 + (36 \p^2 -108) z \right. \right. \nonumber \\
&& \hspace{0.3cm}
-144 \p^2 z^{3/2} + (648 - 648 L + 108 L^2) z^2  \nonumber \\
&& \hspace{0.3cm} \left. \left.
+ (120 \p^2 -314 +12 L +288 L^2) z^3 \right] \right\}
\nonumber \\
&& \hspace{0.3cm} \times
\frac{\a_s}{\p} \, C_F Q_d \bra s \g |O_7| b \ket _{tree}
\eea
\bea
\label{m23reshv}
M_2(3) &=& \left\{ -\frac{1}{8 \e} \,
\left( \frac{m_b}{\mu} \right)^{-4\e} \,
+ \frac{1}{48} \left[ -93/2  \right. \right. \nonumber \\
&& \hspace{0.3cm}
+(72 -12 \pi^2 -96 \zeta (3) + (96 -24 \pi^2) L +12 L^2 + 8L^3) z
\nonumber \\
&& \hspace{0.3cm}
+(60 + 24 \pi^2 -96 \zeta(3) + (24 -24 \pi^2) L -24 L^2 + 8L^3) z^2
\nonumber \\ && \hspace{0.3cm} \left.
-(68-48L) z^3
\right] \nonumber \\
&& \hspace{0.3cm}
+ \frac{i \pi}{12} \left[ -3 +
(24 - 2 \p^2 + 6 L + 6 L^2) z  \right. \nonumber \\ && \hspace{0.3cm}
\left. \left. +
(6-2\pi^2-12 L +6 L^2) z^2 + 12 z^3 \right] \right\}
\nonumber \\
&& \hspace{0.3cm} \times
\frac{\a_s}{\p} \, C_F Q_u \bra s \g |O_7| b \ket _{tree}
\eea
\bea
\label{m24reshv}
M_2(4) &=& \left\{ -\frac{1}{4 \e} \,
\left( \frac{m_b}{\mu} \right)^{-4\e} \,
- \frac{1}{24} \left[ 93/4   \right. \right. \nonumber \\
&& \hspace{0.3cm}
+(-24 + 2 \pi^2 +24 \zeta (3) - (12 -6 \pi^2) L + 2L^3) z
\nonumber \\
&& \hspace{0.3cm}
+(-12- 4\pi^2 -48 \zeta(3) +12 L -6 L^2 + 2L^3) z^2
\nonumber \\ && \hspace{0.3cm} \left. \left.
+(-6+6\pi^2-24L +18 L^2) z^3
\right] \right\}
\nonumber \\
&& \hspace{0.3cm} \times
\frac{\a_s}{\p} \, C_F Q_u \bra s \g |O_7| b \ket _{tree}
\eea
A comparison with
the corresponding NDR scheme expressions in section 2.1
shows that the imaginary parts are identical for each set $M_2(i)$.
This property has to be satisfied because these
imaginary parts
can be derived by cutting rule techniques
where no regularization is necessary. As expected,
the only difference between the two schemes is
in the $m_c^2$ independent terms of the real part.

\subsection{Counterterms}
There are several counterterm contributions (multiplying $C_2$)
which all have to be taken into account. 
As noted, in this section we only give the counterterms whose effects
do not lead to terms linear in $m_c$.
There is just one such counterterm coming from $\delta Z_{27} O_7$.
The operator renormalization
constant $Z_{27}$ \cite{Ciuchini} reads
\be
\label{zfactorhv}
\delta Z_{27} = \frac{\a_s}{\pi \e} \,
(\frac{3}{8} Q_u + \frac{1}{36} Q_d) \, C_F \quad .
\ee
Defining
\be
M_{27} = \bra s \g |\delta Z_{27} O_7|b \ket \quad ,
\ee
the counterterm contribution is given by
\be
\label{counterhv}
M_{27} = \frac{\a_s}{ \pi} \,
\left( \frac{3 Q_u C_F}{8} + \frac{Q_d C_F}{36} \right)
\, \frac{1}{\e} \, \bra s \g |O_7| b \ket _{tree} \quad .
\ee

\subsection{Renormalized contribution proportional to
$C_2$}
Adding the two-loop diagrams (eqs.
(\ref{m21reshv}), (\ref{m22reshv}), (\ref{m23reshv}) and
(\ref{m24reshv}))
and the counterterm
(eq. (\ref{counterhv})),
we arrive at the renormalized contribution
which we denote by $M_2$.
\be
\label{m2formalhv}
M_2 = M_2(1) + M_2(2) + M_2(3) + M_2(4) + M_{27} \quad .
\ee
Inserting $C_F=4/3$, $Q_u=2/3$ and
$Q_d=-1/3$  we get
in the HV scheme
\be
\label{m2lrhv}
M_2 = \bra s \g |O_7| b \ket _{tree} \, \frac{\a_s}{4 \p} \,
\left( \ell_2 \log \frac{m_b}{\mu}  + r_2 \right) \quad ,
\ee
with
\be
\label{l2hv}
\ell_2 = \frac{416}{81}
\ee
\bea
\label{rer2hv}
\Re r_2 &=& \frac{2}{243} \, \left\{- 860 + 144 \pi^2 z^{3/2}
\right. \nonumber \\
&& \hspace{0.3cm}
+ \left[ 1728 -180 \pi^2 -1296 \zeta (3) + (1296-324 \pi^2) L +
108 L^2 + 36 L^3 \right] \, z \nonumber \\
&& \hspace{0.3cm}
+ \left[ 648 + 72 \pi^2 + (432 - 216 \pi^2) L + 36 L^3 \right] \, z^2
\nonumber \\
&& \hspace{0.3cm}        \left. +
\left[ -54 - 84 \pi^2 + 1092 L - 756 L^2 \right] \, z^3 \, \right\}
\eea
\bea
\label{imr2hv}
\Im r_2 &=& \frac{16 \p}{81} \, \left\{- 5
+ \left[ 45-3 \pi^2 + 9 L +
9 L^2 \right] \, z
+ \left[ -3 \pi^2 + 9 L^2 \right] \, z^2 +
\left[ 28 - 12 L  \right] \, z^3 \, \right\}
\eea
Again, $\Re r_2$ and $\Im r_2$ denote the real and the imaginary part
of $r_2$, respectively.
Comparing with the final renormalized expression
$M_2$ in the NDR scheme, given in section 2.3, we conclude 
again that the scheme dependence
only affects the $m_c^2$ independent term in the real part of $r_2$.

\subsection{Terms linear in $m_c$}
Finally, we collect
the linear $m_c$- terms
of the two-loop digrams
which stems from  in the building blocks
in eqs. (\ref{build1hv}) and (\ref{build2hv}). 
We obtain
\be
\label{m21chv}
M_2^c(1) = - \frac{1}{24} \, \left[ \frac{2}{ \e} \,
+17 - 8 \log \frac{m_b}{\mu} +4 \p i \right]
\,
\frac{\a_s}{\p} \, C_F Q_d \bra s \g |O_{7L}^{new}  | b \ket _{tree}
\ee
\be
\label{m22chv}
M_2^c(2) = - \frac{1}{24} \, \left[ \frac{2}{ \e} \,
+11 - 8 \log \frac{m_b}{\mu}  \right]
\,
\frac{\a_s}{\p} \, C_F Q_d \bra s \g |O_{7R}^{new}  | b \ket _{tree}
\ee
\be
\label{m234chv}
M_2^c(3)+M_2^c(4) = - \frac{1}{4} \, \left[ \frac{2}{ \e} \,
+3 - 8 \log \frac{m_c}{\mu}  \right]
\,
\frac{\a_s}{\p} \, C_F Q_u 
\bra s \g |O_7^{new}  | b \ket _{tree}
\quad ,
\ee
where the new-induced magnetic type operator $O_7^{new}$
in eq. (\ref{m234chv}) reads
\be
\label{o7new}
O_7^{new} = m_c(\mu) \, \frac{e}{16\p^2} \, \bar{s} \, \sigma_{\mu \nu}
b \, F^{\mu \nu} \quad .
\ee
The operators $O_{7L}^{new}$ or $O_{7R}^{new}$, which only
enter in intermediate steps, contain an additional left- or right-
handed projection operator after the $\sigma_{\mu\nu}$ term
in eq. (\ref{o7new}).  

In addition to these two-loop diagrams, there are two counterterms
proportional to evanescent operators which
lead to terms linear in  $m_c$. 
The first (denoted by $1/\e \, O_{4\, Fermi}^{ev}$) 
comes from one-loop gluon corrections to the
four-Fermi operator $O_2$ and is given explicitly in ref.
\cite{Buras1990}. As we use this operator only as an insertion
into the matrix
element for $b \to s \g$, we adapt  the general color structure given
in \cite{Buras1990} for this special case.
In  our notation this counterterm takes the form
\be
\label{counterburas}
\frac{1}{\e} O_{4\,Fermi}^{ev} = - \frac{\a_s}{8\pi} \, 
\frac{1}{\e} \, C_F \, 
\mbox{E}_{HV} \quad ,
\ee
with
\bea
\label{burasehv}
\mbox{E}_{HV} &=& -12 \e \left( \G_- \otimes \G_+ +
\G_+ \otimes \G_- \right)
- \left[ \tilde{\g}_{\tau} \tilde{\g}_\rho \hat{\g}_\mu \g_5 \otimes
           \tilde{\g}^{\tau} \tilde{\g}^\rho \hat{\g}^\mu \g_5 +
           2 \hat{\g}_\mu \otimes \hat{\g}^\mu-
           2 \hat{\g}_\mu \g_5 \otimes \hat{\g}^\mu \g_5 \, \right]
\nonumber \\
&& - \left[ \g_\mu \g_\rho \G_- \g^\rho \g^\mu \otimes \G_- -
            \g_\mu \g_\rho \G_- \otimes \G_- \g^\rho \g^\mu -
           \G_- \g_\rho \g_\mu  \otimes  \g^\mu \g^\rho \G_- +
           \G_- \otimes \g_\mu \g_\rho \G_- \g^\rho \g^\mu \, \right]
\quad ,
\nonumber \\
\eea
where $\G_\pm=\tilde{\g}_\mu \, (1 \pm \g_5)/2$.
Its contribution to the
amplitude for $b \to s \g$ is
\be
\label{m2burasev}
M_2^{ev}[\mbox{four-Fermi}] = \left(\frac{1}{\e} - 2 \log \frac{m_c}{\mu} 
\right) \, 
 \frac{\a_s}{\pi} C_F Q_u \,
\bra s \g |O_7^{new}  | b \ket _{tree} \quad .
\ee 

The second counterterm 
(denoted by $1/\e \, O_{penguin}^{ev}$) 
corresponds to the $1/\e$ pole term in
$m_c$-term in the one-loop building block $I_\b$ in eq. (\ref{build1hv}).
It can be written as
\be
\label{conterpeng}
\frac{1}{\e} \, O_{penguin}^{ev} \,
=  \frac{1}{\e} \, \frac{g_s}{16\pi^2} \, m_c \,
\bar{s} \, R \, \sigma_{\mu \nu} L \frac{\l^A}{2} \, b \, \, 
G_A^{\mu \nu} \quad .
\ee
Its contribution to the amplitude $b \to s \gamma$
is given by 6 graphs whose Feynman diagrams are similar
to those shown in Fig. \ref{fig:8}.
The 3 diagrams, where the gluon is absorbed by the $s$-quark,
read
\be
\label{diapenga}
M_2^{ev}(penguin \,a) = \frac{1}{36} \, \left[ \frac{6}{ \e} \,
+19 - 12 \log \frac{m_b}{\mu} +6 \p i \right]
\,
\frac{\a_s}{\p} \, C_F Q_d \bra s \g |O_{7L}^{new}  | b \ket _{tree}
\quad , 
\ee
while the diagrams where the gluon is absorbed by the $b$-quark 
is given by
\be
\label{diapengb}
M_2^{ev}(penguin \,b) = \frac{1}{18} \, \left[ \frac{3}{ \e} \,
+5 - 6 \log \frac{m_b}{\mu}  \right]
\,
\frac{\a_s}{\p} \, C_F Q_d 
\bra s \g |O_{7R}^{new} | b \ket _{tree}
\quad . 
\ee

Adding the contributions of the two-loop diagrams in eqs. 
(\ref{m21chv}, \ref{m22chv}, \ref{m234chv}) 
and the counterterm contributions in eqs.
(\ref{m2burasev},\ref{diapenga},\ref{diapengb}),
we end up with a total linear $m_c$  term of the form
\be
\label{m2cfinal}
M_2^c[two \, loop \, + \, countert.] = \frac{\a_s}{\pi} \, C_F \, \left[ 
\frac{Q_d}{72} \left(  \frac{6}{\e} - 13 \right)
+ \frac{Q_u}{4} \left(  \frac{2}{\e} - 3 \right) 
\right] \,
\bra s \g |O_{7}^{new} | b \ket _{tree}
 \quad.
\ee  

Of course, this result can be made finite by introducing
a corresponding counterterm proportional to $O_7^{new}$ given in eq.
(\ref{o7new}) which minimally subtracts the $1/\e$-pole in eq. 
(\ref{m2cfinal}).

At this point one might ask why these additional operators
appearing in this subsection do not appear in 
the literature
\cite{Ciuchini} 
where the singularity structure of these graphs 
have been worked out in order to
extract the $O(\a_s)$ anomalous dimension matrix.
We also note that there are other new operators induced if one
looks e.g. at the analogous two-loop corrections to $b \to s \gamma$
associated to other Four-Fermi operators. For example, the
corrections to $O_3$ generate a new magnetic type operator,
where $m_c$ in eq. (\ref{o7new}) is replaced by $m_b$.

To absorb all the divergencies one clearly has to enlarge the
operator basis in this scheme. However, we believe that it is
correct to ignore these additional operators for the
leading logarithmic result for $b \to s \gamma$.   
The reason is the following: First, the new operators do not mix into
the old ones given in eq. (\ref{operators}) at $O(\a_s)$;
therefore, the old operators run in the same way with or without
including the new operators.
Second, in the $O(\a_s^0)$ matching (at $\mu=m_W$), the new
coefficients have to be zero, as one can easily see. 
Third, the absence of a $\log(\mu)$ term in eq. (\ref{m2cfinal})
indicates, that the four-Fermi operators do not induce any
running of the new
operators at the leading logarithmic level. 
The only way, the new magnetic operators could run
is by multiplicative renormalization. But as these operators
have zero initial value as discussed above, 
also this effect is unimportant: To leading logarithmic 
precision the Wilson coefficients  of the
new operators are zero at each renormalization scale.
Therefore the new operators certainly do not change the leading
logarithmic physics. Equivalently, one can say that one 
can throw  away the terms proportional to $m_c$ in
the building blocks in eqs. (\ref{build1hv}) and (\ref{build2hv})
when working at leading order.

However, our result in eq. (\ref{m2cfinal}) shows that 
the matrix element of $O_2$ leads to finite terms 
(which are of next-to-leading
order) linear in $m_c$ when calculated in the HV scheme. 
It would be very interesting to see,
if these  linear  $m_c$ term are cancelled by the next-to-leading
Wilson coefficients of the new operators. 
Such a cancellation is expected to occur, of course,
because in any other scheme which respects the 4-dimensional
chirality properties, these terms do not appear.      

\section{Bremsstrahlung corrections}
\setcounter{equation}{0}
In order to make the paper self-contained, we give in this appendix 
the formulae for the process $b \to s \gamma g$ based on the operators 
$O_2$, $O_7$ and $O_8$
in the NDR scheme. As we have given the analogous virtual 
corrections to $b \to s \gamma$ 
in the limit $m_s \to 0$, we also present the 
Bremsstrahlung corrections for this case.

We denote the various contributions to the 
Bremsstrahlung decay width 
$\Gamma^b$ by $\Gamma^b_{22}$, 
$\Gamma^b_{77}$, $\Gamma^b_{88}$, $\Gamma^b_{27}$, $\Gamma^b_{28}$, 
$\Gamma^b_{78}$; for example, $\Gamma^b_{22}$ is 
based on the 
matrix element squared of $O_2$, while $\Gamma^b_{27}$ is an 
interference term between the matrix element of $O_2$ and $O_7$, etc. .

Note that all the interference terms and $\Gamma^b_{22}$ are 
infrared (and collinear) finite. 
The sum of these four finite contributions is given by
\be
\label{bre1}
 \Gamma_F^{brems} = \frac{G^2_F | \lambda_{t} |^2 
\alpha_{em} \alpha_{s}}{768 \pi^5 m_b} 
\int_{PS}  dE_g dE_{\gamma} \,
(\tau_{22}+\tau_{27}+\tau_{28}+\tau_{78}) 
\ee
\be
\label{bre2}
\tau_{22}=2m^2_b \, Q^2_u \, C^2_2(\mu) \, |\kappa|^2 \,
(m_b^2 -2 (qr))  
\ee
\be
\label{bre3}
\tau_{27}=-32 m_b^2 \, Q_u \, C_2(\mu)  C_7^{eff}(\mu) \, (qr) \,
Re(\kappa)
\ee
\be
\label{bre4}
\tau_{28}=-32 m_b^2 \, Q_u Q_d \, C_2(\mu) C_8^{eff} \, (qr) \,
Re(\kappa)
\ee
\be
\label{bre5}
\tau_{78}=-128 m_b^2 \, Q_d \, C_7^{eff}(\mu)  C_8^{eff}(\mu) \,
(qr) \, 
\frac{m_b^4+2(pq)(pr)}{(pq)(pr)} \quad ,
\ee
where as in the previous sections $p$, $p'$, $q$ and $r$ 
denote the four-momenta of the $b$- and the $s$-quark, 
the photon and the gluon respectively. The function $\kappa$
is defined as
\be
\label{bre6}
\kappa  = \frac{4(2G(t)+t)}{t} \quad ; \quad t=\frac{2(qr)}{m_c^2}
\ee
\be
\label{bre7}
G(t)= \int_0^1\frac{dy}{y} \log \left[ 1-ty(1-y)-i\epsilon \right]
\quad .
\ee
The phase space boundaries (denoted above by $PS$) are given by 
\be
\label{bre8}
E_\g \in [ 0, \frac{m_b}{2} ]; \quad E_g \in [ \frac{m_b}{2}-E_\g,
 \frac{m_b}{2}]. 
\ee
Note, that the function $\kappa$ in eq.
(\ref{bre6}) is finite 
for $m_c \to 0$. Therefore also the Bremsstrahlung 
corrections have a finite limit for $m_c \to 0$.

$\Gamma_{77}^b $ is singular for $E_g \to 0$ or 
$\vec{r}\parallel \vec{p}'$. As it cancels the corresponding 
singularity of the virtual corrections to $O_7$, we
have taken into account the contribution $\Gamma^b_{77}$ 
already in section 3, i.e., $\Gamma^b_{77}$ is contained in the 
finite quantity in eq. (\ref{Gamma7}).

Finally, $ \Gamma^b_{88} $ is singular for $E_\gamma \to 0$ or 
$\vec{q} \parallel 
\vec{p}'$. These singularities can be removed by adding the virtual photon 
corrections to $ b \to s g $. This finite sum $ \Gamma_{88} $ can easily 
be obtained from $ \Gamma_{77} $ in eq. (\ref{Gamma7}). It reads 
\be
\label{bre9}
\Gamma_{88}=\frac{m_b^5}{96 \pi^5} |G_F Q_d C_8^{eff} 
\lambda_t|^2 \alpha_{em} \alpha_s \left( \frac{16}{3}-
\frac{4\pi^2}{3}+ 4 \log \frac{m_b}{\mu} \right) \quad .
\ee
To summarize, the total inclusive decay width $\Gamma$ for 
$\BGAMAXS$ is then 
given by $\Gamma = \Gamma^{virt} + \Gamma^{brems}_F + \Gamma_{88} $,   
where $\Gamma^{virt}$, $\Gamma^{brems}_F$ and   
$\Gamma_{88}$ are given in eqs. 
(\ref{widthvirt}), (\ref{bre1}) and (\ref{bre9}), respectively.   


\end{document}